\documentclass{article}

\usepackage{arxiv}

\usepackage[utf8]{inputenc}
\usepackage{amsmath,amssymb}
\usepackage[hidelinks]{hyperref}%
\usepackage{url}%
\usepackage{geometry}
\usepackage{graphicx}
\usepackage{tikz,pgfplots}
\usetikzlibrary{arrows}
\usepgfplotslibrary{groupplots}
\usepackage{bm}
\usepackage{algorithm}
\usepackage{algorithmicx}
\usepackage{algpseudocode}
\usepackage{booktabs}
\usepackage{doi}%
\usepackage{subcaption}%
\usepackage{multirow}%
\usepackage{comment}
\RequirePackage[authoryear]{natbib}
\usepackage{cuted}

\algdef{SE}[DOWHILE]{Do}{doWhile}{\algorithmicdo}[1]{\algorithmicwhile\ #1}%

\title{A Gaussian Sliding Windows Regression Model for Hydrological Inference}
\renewcommand{\shorttitle}{A Gaussian Sliding Windows Regression Model}

\author{\href{https://orcid.org/0000-0003-1327-4855}{\includegraphics[scale=0.06]{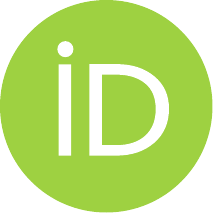}\hspace{1mm}
Stefan Schrunner}\thanks{Current affiliation: Carinthia University of Applied Sciences, Villach, Austria}\\
Department of Data Science\\
Norwegian University of Life Sciences\\
Ås, Norway\\
\texttt{stefan.schrunner@fh-kaernten.at}\\
\And
\href{https://orcid.org/0009-0003-4422-8111}{\includegraphics[scale=0.06]{orcid.pdf}\hspace{1mm}
Parham Pishrobat}\\
Department of Statistics\\
University of British Columbia\\
Vancouver, Canada.\\
\texttt{par.pishrobat@ubc.ca}\\
\And
\href{https://orcid.org/0000-0003-4200-7895}{\includegraphics[scale=0.06]{orcid.pdf}\hspace{1mm}
Joseph Janssen}\\
Department of Earth, Ocean and Atmospheric Sciences\\
University of British Columbia\\
Vancouver, Canada\\
\texttt{joejanssen@eoas.ubc.ca}
\And
\href{https://orcid.org/0000-0002-6919-3483}{\includegraphics[scale=0.06]{orcid.pdf}\hspace{1mm}
Anna Jenul}\\
Department of Data Science\\
Norwegian University of Life Sciences\\
Ås, Norway\\
\texttt{anna.jenul@gmail.com}\\
\And
\href{https://orcid.org/0000-0001-7417-6330}{\includegraphics[scale=0.06]{orcid.pdf}\hspace{1mm}
Jiguo Cao}\\
Department of Statistics and Actuarial Science\\
Simon Fraser University\\
Burnaby, Canada.\\
\texttt{jiguo\_cao@sfu.ca}\\
\And
\href{https://orcid.org/0000-0002-8173-887X}{\includegraphics[scale=0.06]{orcid.pdf}\hspace{1mm}
Ali A. Ameli}\\
Department of Earth, Ocean and Atmospheric Sciences\\
University of British Columbia\\
Vancouver, Canada.\\
\texttt{aameli@eoas.ubc.ca}\\
\And
\href{https://orcid.org/0000-0002-4575-3124}{\includegraphics[scale=0.06]{orcid.pdf}\hspace{1mm}
William J. Welch}\\
Department of Statistics\\
University of British Columbia\\
Vancouver, Canada.\\
\texttt{will@stat.ubc.ca}\\
}

\begin{document}

\maketitle

\begin{abstract}
    Statistical models are an essential tool to model, forecast and understand the hydrological processes in watersheds. In particular, the understanding of time lags associated with the delay between rainfall occurrence and subsequent changes in streamflow is of high practical importance. Since water can take a variety of flow paths to generate streamflow, a series of distinct runoff pulses may combine to create the observed streamflow time series. Current state-of-the-art models are not able to sufficiently confront the problem complexity with interpretable parametrization, thus preventing novel insights about the dynamics of distinct flow paths from being formed. The proposed Gaussian Sliding Windows Regression Model targets this problem by combining the concept of multiple windows sliding along the time axis with multiple linear regression. The window kernels, which indicate the weights applied to different time lags, are implemented via Gaussian-shaped kernels. As a result, straightforward process inference can be achieved since each window can represent one flow path. Experiments on simulated and real-world scenarios underline that the proposed model achieves accurate parameter estimates and competitive predictive performance, while fostering explainable and interpretable hydrological modeling.
\end{abstract}

\section{Introduction}
The hydrological processes that produce streamflow play key roles in determining the environmental effects of climate and land use changes. In particular, changes in climate or land use can trigger a complex series of nonlinear and interactive processes which can eventually impact the way in which watersheds partition, store, and release water, leading to potential changes in flood, landslide or drought risks \citep{dunn2010transit,harman2011functional,sawicz2014characterizing}. Furthermore, water reservoirs and consistent streamflow are essential for the regulation of the water supply in both urban and rural areas \citep{janssen2021assessment,tang2009use}, as well as for sustainable energy production and continually healthy ecological habitats \citep{ahmad2020maximizing,zalewski2000ecohydrology}. A deep understanding of how precipitated water becomes streamflow is of great importance and requires a sophisticated and interpretable statistical framework.

Hence we investigate the problem of inferring streamflow partitioning into different flow paths such as overland flow, subsurface flow, and baseflow \citep{cornette2022hillslope,kannan2007hydrological,mcmillan2020linking,nejadhashemi2009case,wang2024towards}. While there exist many simple methods which can partition streamflow into fast surface flow versus slow baseflow \citep{nejadhashemi2009case,wang2024towards}, the processes and reasoning behind this partitioning are usually lost. To prevent this, one could consider how input rainfall is partitioned into different flow paths. Hence, we consider a time series $\bm{y}_T =\left(y_t\right)_{t\in T}$
of streamflow at a particular gauge and an input time series $\bm{x}_T =\left(x_t\right)_{t\in T}$ 
of rainfall data on a common time domain $T$. Given such data, a common approach is a lagged regression model following the concepts of auto-regressive moving-average with exogenous variables (ARIMAX, ARMAX) \citep{hyndman2021arima}. In contrast to these models, our problem does not involve an auto-regressive part, i.e., we do not use lagged observations of the target $y_t$ as predictors; instead, we seek to fully characterize the target time series via the lagged input time series. Further, we desire more interpretability and inferential power than is offered by ARIMAX models. For example, the maximum time interval for which the predictor has an impact on the response and the separation of flow pathways is of interest for watershed modeling.

A more closely related model in time series analysis is the finite distributed lag model (DLM)~\citep{baltagi2022econometrics}. Despite its roots in econometrics, the DLM has been extensively applied in environmental sciences~\citep{chen2018dlmenvironmental,peng2009dlmpollution,rushworth2013distributed,warren2020dlmpollution}. While ARIMAX mainly relies on the concept of ARIMA, i.e. it makes use of the lagged target variable $y_t$ as input, and only adds the present time point of the exogenous variable $x_t$, DLM predicts the target variable purely based on a distinct lagged input time series --- the same setup that is pursued in this work. In this sense, the presented work may be seen as specific type of a DLM and thereby an extension of the work by \cite{rushworth2013distributed}, however, with an improved parameterization, which allows for improved model interpretability. Basic DLM variants do not allow us to impose parametric assumptions on the shape of the regression parameter vector beyond simple singular shapes like geometrically decreasing patterns or polynomials \citep{almon1965distributed,eisner1960distributed,griliches1967distributed}, though recent work \citep{rushworth2018bayesian} extends the concept to B-splines in a Bayesian setting. Such DLM extensions permit a large variety of lag curve shapes, but do not make the separation between flow paths explicit. To the best of our knowledge, none of the existing DLM variants is capable of explicitly separating distinct flow pulses, which is most crucial for hydrologic inference.

There are several other possible approaches.
Another economic model was adapted for hydrological inference by \cite{giani2020}, where the authors deploy the concept of cross-correlations in combination with a moving-average approach. Although their method has favourable properties to model the response time between rainfall events and flow pulses, it again does not allow a direct separation of the flow paths. Furthermore, correlation-based analyses do not allow straightforward predictions on new test datasets.
Similarly, we do not pursue continuous-time models in hydrology based on partial differential equations \citep[e.g.][]{young2006transfer} or functional data analysis \citep[e.g.][]{janssen2023learning}. 

In contrast to further related works in the field of statistical modeling in hydrology, the methodology presented here focuses exclusively on the temporal relation between rainfall and streamflow in a given watershed. Spatial relationships between watersheds, as investigated by \cite{roksvg2021runoff}, are beyond the scope of this article; nevertheless, the presented concept offers a foundation, which can be extended to include spatial information for gauged and ungauged catchments in future work and may complement existing large-scale analyses, such as that of \cite{hare2021continentalscale}.

In this article, we present a parameterized, interpretable variant of lagged regression models for hydrological modelling: the Gaussian Sliding Windows Regression model. It builds on the assumption that distinct flow paths (surface flow, sub-surface flow, etc.) provoke separable pulses of streamflow after rainfall events. These pulses are represented by multiple temporal windows, which weight the input $x_t$ at time points within the window via a Gaussian kernel. The output of each window is mapped to the target variable $y_t$ using a linear regression model. Compared to earlier Sliding Windows Regression models \citep{davtyan2020slidingwindow,khan2019slidingwindow,janssen2021assessment}, the proposed method has major structural differences, which constitute the main novelties of our work:
\begin{enumerate}
    \item We consider multiple, potentially overlapping windows representing distinct flow paths, and
    \item we use a parameterized kernel as model weights to allow straightforward interpretations for hydrological inference of flow path importance.
\end{enumerate}
Our experimental evaluation involves both simulated and real-world data. The simulation study serves as a proof of concept for the model and validates the parameter estimation procedure. The application to two real-world datasets demonstrates that the model achieves competitive predictive performance and has favourable properties for interpreting underlying hydrological processes.

\section{Sliding Windows Regression Model}
We assume that a univariate input time series $\bm{x}_T$ and a univariate target time series $\bm{y}_T$ are given on a common, discrete time domain $T = [t_{\max}]$, where $[n]$ denotes the index set $\{0,\dots,n\}$. In the motivating application $\bm{x}_T$ and $\bm{y}_T$ represent daily rainfall and streamflow, respectively. The modeling goal is to represent the target value $y_t$ at time index $t\in T$ based on time-lagged observations of $\bm{x}_{[t]}$, 
i.e., up to time $t$.

Our proposed model pursues the idea that the effect of precipitation on streamflow follows a mixture of $k$ kernels associated with temporal windows, each one characterizing a particular flow path (groundwater flow, sub-surface flow, or overland flow). Ultimately, we aim to represent the distribution of response times from flow path $i\in \{1,\dots,k\}$ by weight vector $\bm{\kappa}^{(i)} \in \mathbb{R}^{t+1}$. Note that the length of $\bm{\kappa}^{(i)}$ is formally time-dependent based on its dimension. Since the sub-vector of non-zero weights remains the same over time, zero padding is used at the end to comply with the dimensionality of $\bm{x}_{[t]}$. Thus, the time index is left out in a slight abuse of notation. Computing a weighted sum of the lagged time series $\bm{x}_{[t]}$ with one such weight vector $\bm{\kappa}^{(i)}$ models the contribution to streamflow $y_t$ associated with flow path $i$. Summing the contributions gives the overall flow. To facilitate explanations, we present the model for the special case of only $k=1$ window first, and extend it to multi-window scenarios afterwards.

\subsection{Single-window model}

If only one flow path exists, we denote the associated weight vector by $\bm{\kappa}\in\mathbb{R}^{t+1}$. By default, we require that all entries in $\bm{\kappa}$ are non-negative, and that $\bm{\kappa}$ is normalized, i.e. $\Vert \bm{\kappa}\Vert_1=1$. Then, in the special case of one window, our \textit{Sliding Windows Regression (SWR)} model describes the gauged runoff $y_t$ by the weighted sum or the associated convolution
\begin{align}
    y_t &= \beta \cdot \left(\sum\limits_{s=0}^t x_{t-s} \kappa_s\right) + \varepsilon_t \nonumber \\
    &= \beta \cdot \left({\bm x}_{[t]} \ast \bm{\kappa}\right) + \varepsilon_t,
    \label{eq:weighted_sum}
\end{align} 
with some multiplicative constant $\beta$ and an error term $\varepsilon_t$ assumed as Gaussian white noise. The symbol $\ast$ denotes the discrete convolution operator defined as $$ \bm{v}\ast\bm{z} = \sum\limits_{i=0}^{n} v_{n-i}z_i$$ for two vectors $\bm{v},\bm{z}\in\mathbb{R}^{n+1}$. Note that weight $\kappa_s$ is applied to lag $s \in [t]$, i.e. $\kappa_s$ denotes the weight applied to observation $x_{t-s}$ rather than $x_s$. The model parameter $\beta$ acts as a regression coefficient to adjust the scaling of the normalized weight vector $\bm{\kappa}$. Since additional precipitation cannot lead to decreased streamflow, we further require $\beta > 0$. In other applications, this restriction may be modified.

Unlike typical linear regression models, the SWR model is defined without an intercept $\beta^{(0)}$. This is due to the assumption that runoff $y_t$ is $0$ after a long period without precipitation, $x_s = 0$ for all $s<t$. However, in other applications, the proposed SWR model may be used with an intercept. 

Further, note that the model definition allows negative responses $y_t$ if all entries of $\beta$ are very small while the error $\varepsilon_t$ takes a larger negative value. However, in this rare case, point forecasts remain non-negative due to non-negative observations $\bm{x}_{[t]}$.

The assumption of independent errors $\varepsilon_t$ in model~(\ref{eq:weighted_sum}) (or its multi-kernel generalization) is unrealistic for most real-world datasets, where autocorrelated disturbances are common. Later, we describe a simple data transformation to address that complexity, however, it is sufficient to proceed with the simpler model~(\ref{eq:weighted_sum}) for now. 

The proposed SWR model~(\ref{eq:weighted_sum}) is related to a distributed lag model, which is a common approach also used in related work \citep{rushworth2013distributed}. However, the paradigm where each flow path is represented by one kernel $\bm{\kappa}$ allows us to make further assumptions to parametrize the lag vector: We assume that each flow path is associated with a distinct unimodal distribution of the time lag (time delay) between a rainfall event and its effect on streamflow. Hence, the weight vector $\bm{\kappa}$ represents the kernel of a probability distribution. Further, we denote the time lag with the highest amount of runoff water, represented by the position of the peak of $\bm{\kappa}$ (mode), as a distribution centre parameter $\delta$, where $0 < \delta < t$.

More specifically, we parametrize $\bm{\kappa}$ by a Gaussian-shaped kernel centred around a lag of $\delta$, assuming that the time delay between the rainfall event and the resulting change in the streamflow is approximately Gaussian distributed. 
Since the time series is only measured at discrete time points, the Gaussian kernel is discretized by applying a step function, such that
\begin{equation*}
    \kappa_s = \frac{1}{C} \int\limits_{s-\frac{1}{2}}^{s+\frac{1}{2}} \phi(\tau;\delta,\sigma^2) \ d\tau,
\end{equation*}
where $\phi(.,\delta,\sigma^2)$ denotes the probability density function of a Gaussian distribution with mean $\delta$ and variance $\sigma^2>0$, the kernel indices (lags) $s\in\{0, 1,\dots,t\}$ are relative to a current time point, and $C$ denotes the normalization constant
\begin{equation}
    C = \sum\limits_{s=0}^{t}\int\limits_{s-\frac{1}{2}}^{s+\frac{1}{2}} \phi(\tau;\delta,\sigma^2) \, d\tau.
    \label{eq:kernel_scaling}
\end{equation} 
Note that the first kernel element, $\kappa_0$, corresponding to lag $s=0$ is supported on $(-\frac{1}{2},\frac{1}{2})$, i.e., including half a time unit before time $t$, because of the discretization.  

The notion of a \textit{temporal window} describes an interval of time lags $\delta \pm r$ with some window size $2r$ centred around $\delta$, which covers a large proportion of the probability mass. Due to the shape of the Gaussian probability density function and the decay when moving away from $\delta$, we define the window such that it covers approximately a range of $\pm$ 3 standard deviations, i.e. $r = 3\sigma$. For now, assume that $\delta - r > 0$ and $\delta + r < t_{\max}$ 
(the special cases where the boundaries of $[t]$ are exceeded will be treated at a later point). Given such a window, the modeled response $y_t$ exclusively depends on 
$\bm{x}_{[t]}$ falling in $(\lfloor t-\delta - r \rfloor, \lceil t-\delta +r \rceil)$ rather than the full index set $[t]$.
Here, $\lfloor \cdot \rfloor$ and $\lceil \cdot \rceil$ denote floor and ceiling operators to consider the weights of the two discrete time points just outside the kernel.
Nevertheless, to keep the notation manageable, we formally retain the notation of $\bm{\kappa}$ covering the whole interval $[t]$ in the following.

Examples for $r\in\{1,3,6\}$ and $\sigma = r / 3$ are depicted in Fig.~\ref{fig:kernel}. Further, the graphical idea behind the SWR model for a univariate input $\bm{x}$ 
can be visualized in Fig.~\ref{fig:single-window}: the target time series $y_t$ at time is predicted based on information originating mainly from the input time series in window $\delta \pm r$.

\begin{figure*}[ht]
    \centering
    \begin{tikzpicture}[scale=0.75]
        \draw[<->] (-8,-0.7) -- (8,-0.7) node[right] {\scriptsize time};
        \node at (7,-1.1) {\tiny{$t$}};
        \foreach \x in {2,4,...,12, 14}
            \node at (7-\x, -1.1) {\tiny{$t-\x$}};
        \foreach \i in {-7,...,7} \draw (\i,-0.7)--(\i,-0.85);
        \draw[fill=yellow, dashed] (-1.3,2.5) rectangle ++(2,3);
        \node at (2.2, 4.75) {$r=1$};
        \draw[domain=-1.3:0.7, smooth, samples = 100] plot ({\x}, {2.5 + exp(-9 * (\x+0.3) * (\x+0.3) / 2) * 3});
        \foreach \x in {-2, -1,0, 1}
            \node at (\x, 2.5) {\textbullet};
        \draw[<->] (0.5,2.3) -- (1.5,2.3);
        \node at (2.5, 2.3) {\tiny{$t-6 \pm 1/2$}};
        \draw[<->] (-2.5,2.3) -- (-1.5,2.3);
        \node at (-3.5, 2.3) {\tiny{$t-9 \pm 1/2$}};
        \draw[fill=yellow, dashed] (-3.3,1) rectangle ++(6,1);
        \node at (4.2, 1.75) {$r=3$};
        \draw[domain=-3:3, smooth, samples = 100] plot ({\x}, {1 + exp(- (\x+0.3) * (\x+0.3) / 2)});
        \foreach \x in {-4,...,3}
            \node at (\x, 1) {\textbullet};
        \draw[<->] (2.5,0.8) -- (3.5,0.8);
        \node at (4.5, 0.8) {\tiny{$t-4 \pm 1/2$}};
        \draw[<->] (-4.5,0.8) -- (-3.5,0.8);
        \node at (-5.5, 0.8) {\tiny{$t-11 \pm 1/2$}};
        \draw[fill=yellow, dashed] (-6.3,0) rectangle ++(12,1/2);
        \node at (7.2, 0.375) {$r=6$};
        \draw[domain=-6:6, smooth, samples = 100] plot ({\x}, {exp(- (\x+0.3) * (\x+0.3) / 8) / 2});
        \foreach \x in {-7,...,6}
            \node at (\x, 0) {\textbullet};
        \draw[<->] (5.5,-0.2) -- (6.5,-0.2);
        \node at (7.5, -0.2) {\tiny{$t-1 \pm 1/2$}};
        \draw[<->] (-7.5,-0.2) -- (-6.5,-0.2);
        \node at (-8.5, -0.2) {\tiny{$t-14 \pm 1/2$}};
        \draw[-, dotted] (-0.3,0) -- (-0.3, 5.75);
        \node at (-0.3, -0.25) {\tiny{$t - \delta$}};
    \end{tikzpicture}
    \caption{Examples of Gaussian shaped window kernels centred around $t-\delta$ with $\delta = 7.3$. For different values of $\sigma$ a window extends approximately $\pm r$ around $t-\delta$, where $r = 3\sigma$.  Each bullet denotes a discrete time point of lag $s$ with weight given by the area under the kernel between $t - s - 1/2$ and $t + s + 1/2$.  A lagged time point at $t -s$ outside $t-\delta \pm r$, such as those at $t-6$ (top), $t-4$ (middle), and $t - 1$ (bottom), may still have a small weight as $t - s \pm 1/2$ falls within the kernel.
    Other time points immediately outside their respective kernels, such as those at $t-9$ (top), $t-11$ (middle), and $t - 14$ (bottom), will be considered but have essentially zero weight as $t - s \pm 1/2$ does not reach non-trivial values of the Gaussian kernel.  
    \label{fig:kernel}}
\end{figure*}

\begin{figure*}[ht]
    \centering
    \begin{tikzpicture}
        \begin{groupplot}[group style={group size=1 by 2},xmin=0,ymin=0,xmax=90,height=4cm,width=\textwidth,no markers]
            \nextgroupplot[xtick={0,20,40,60,80,90},
                           xticklabels = {0,20,40,60,80,$t$},
                          ytick={0,5,10,15},
                          yticklabels={0,5,10,$y_t$~~},
                          ymax = 15]
            \addplot table [x=t, y=x, col sep=comma] {resp.csv};
            \node (a1) at (70, 10) {\textbullet};
            \draw[<->] (70, 50) -- (50, 50);
            \node at (60, 60) {$\delta$};
            \draw[dotted] (50,50) -- (50,0);
            \draw[dotted] (70,50) -- (70,0);
            \nextgroupplot[xtick={0,20,40,50,60,70,80,90},
                          xticklabels={0,20,40,$s - \delta$,60,$s$,80,$t$},
                          ytick={0,20,40,60,80},
                          yticklabels={0,20,40,60,$x_t$~~},
                          ymax = 80]
            \draw[dashed,fill=yellow] (40, 0) rectangle ++(20,300);
            \addplot table [x=t, y=x, col sep=comma] {rain.csv};
            \draw[<->] (40, 400) -- (50, 400);
            \node at (45, 500) {$r$};
            \node (b1) at (60, 300) {};
            \draw[dotted] (50,0) -- (50,1000);
            \draw[dotted] (70,0) -- (70,1000);
        \end{groupplot}
        \draw[->] (b1) to [out=60,in=270] (a1);
    \end{tikzpicture}
    \caption{Illustration of the Sliding Windows Regression model with $k=1$ window, which predicts $y_{70}$ based on a subset of ${\bm x}_{[70]}$ values in a window. With $\delta = 20$ and $ r =10$, the window $[40,60]$ centred around $s-\delta=50$ covers $2r + 1 = 21$ time points.}
    \label{fig:single-window}
\end{figure*}

In the special case that $\sigma$ is very small and $r$ approaches $0$, the window width collapses to $0$, and the full probability mass of the kernel accumulates at the window centre $\delta$. However, since $\delta$ may be located between two observations of the input time series, $\sigma$ is restricted to be at least $\frac{1}{6}$ to ensure at least one observation falls within the effective domain of the probability mass, i.e., is assigned to the closest measured time point in such cases, by definition.

Non-zero weights can only be assigned to lags in $[t]$, and it follows that the Gaussian SWR model cannot be evaluated for arbitrary parameter combinations of $\delta$ and $\sigma$. The restriction on the upper (lag) limit of the window, i.e., $\delta + r \leq t$, is less prohibitive in practice for longer time series and merely leads to $y_t$ not being predictable for $t < \delta + r$, i.e. for small $t$ at the beginning of the time series. As a consequence, this restriction can be easily resolved by withholding data for calibration during model training. Hydrological applications typically have time lags $\delta$ in the range of a few days, weeks or months, hence the impact of this restriction is negligible for a sufficiently long time series spanning over multiple years.

On the other hand, peaks appearing at short time lags are highly relevant in practice, leading to small values of $\delta$. Under extremely wet conditions in low permeability catchments, it can be assumed that water from rainfall will almost immediately lead to changes in streamflow. If the lower limit of the lag window, $\delta - r$, falls below $0$, we change the shape of the kernel to a truncated Gaussian kernel, as shown in Figure \ref{fig:trunckernel}. Note that the kernel is, by definition in (\ref{eq:kernel_scaling}) always normalized.

\begin{figure*}[ht]
    \centering
    \begin{tikzpicture}[scale=0.75]
        \draw[<->] (-7,-0.5) -- (7,-0.5) node[right] {\scriptsize time};
        \fill[fill=yellow] (-5.7,0) rectangle ++(9.2,0.5);
        \draw[dashed] (-5.7,0) rectangle ++(12,0.5);
        \node at (7.5, 0.25) {$r=6$};
        \node at (0.3,-0.25) {\tiny{$t-\delta \pm r$}};
        \draw[dotted][->] (1.3,-0.25) -- (6.3,-0.25);
        \draw[dotted][<-] (-5.7,-0.25) -- (-0.7,-0.25);
        %
        \draw[domain=-6:3.5, smooth, samples = 100] plot ({\x}, {exp(- (\x-0.3) * (\x-0.3) / 8) / 2});
        \draw[-, dotted] (0.3,0) -- (0.3, 0.5);
        %
        \foreach \x in {-6,...,6}
            \node at (\x, 0) {\textbullet};
        \foreach \i in {-6,...,6} \draw (\i,-0.5)--(\i,-0.65);
        \foreach \x in {1,...,9}
            \node at (3-\x,-0.9) {\tiny{$t - \x$}};
        \node at (3, -0.9) {\tiny{$t$}};
        \draw[<->] (2.5,1) -- (3.48,1);
        \node at (3,1.5) {$t\pm \frac{1}{2}$};
        \draw[<->] (3.52,1) -- (6.3,1);
        \node at (4.5, 1.5) {$\tau$};
    \end{tikzpicture}
    \caption{Example of a truncated Gaussian shaped window kernel, with window parameters $\delta = 2.7$ and $r=3\sigma=6$.  The window spanning $t - \delta \pm r$ shown by the dashed rectangle is truncated to the shaded (yellow) region: rainfall after time $t$ cannot contribute to $y_t$. Hence, the weights for $x_{t+1}$, $x_{t+2}$, and $x_{t+3}$ in the time interval marked $\tau$ are all zero even though the interval is within the rectangle. The Gaussian curve is renormalized to 1. The kernel weight applied to $x_t$ is the area under the curve between $t - \frac{1}{2}$ and $t + \frac{1}{2}$ due to discretization.  }
    \label{fig:trunckernel}
\end{figure*}

Truncation has effects on the interpretation of the model parameters: while $\delta$ still indicates the time lag with peak kernel weight (mode of the probability distribution), it does not indicate the central lag of the window on the time axis anymore. Further, the right-truncated Gaussian distribution is no longer symmetric with respect to the temporal window, and hence, the distribution mean shifts to the left.

\subsection{Multi-window model}
Assuming multiple different (potentially overlapping) windows allows the model to account for distinct effects on the current target $y_t$. For this purpose, we extend our method from a single-window model to a mixture of $k$ kernels, which is related to the density resulting from a mixture of probability distributions~\citep{everitt1981gmm}. For a given window $i\in \{1,\dots,k\}$, we use the same notation as in the single-window case for the window and kernel parameters with an added index $i$ denoting the window number: $\delta^{(i)}$, $r^{(i)}$, and $\sigma^{(i)}$. The Gaussian SWR model with multiple windows is given as a direct generalization of (\ref{eq:weighted_sum}) by
\begin{align}
    y_t &= \sum_{i=1}^k \beta^{(i)} \cdot \left({\bm x}_{[t]} \ast \bm{\kappa}^{(i)}\right) + \varepsilon_t, \\
    &= {\bm x}_{[t]} \ast \left(\sum\limits_{i=1}^k \beta^{(i)} \bm{\kappa}^{(i)}\right) + \varepsilon_t
    \label{eq:multiwindow}
\end{align}
where the regression parameters $\beta^{(1)},\dots,\beta^{(k)} \geq 0$ act as window weights.
In summary, the Gaussian SWR model with $k$ windows contains the following parameters
\begin{itemize}
    \item $k$ regression parameters $\bm{\beta} = (\beta^{(1)},\dots,\beta^{(k)}$), $\beta^{(i)}\in\mathbb{R}^+$,
    \item $k$ lag parameters $\bm{\delta} = (\delta^{(1)},\dots,\delta^{(k)}$), $\delta^{(i)}\in \mathbb{R}^+$, and
    \item $k$ kernel standard deviation (size) parameters $\bm{\sigma} = (\sigma^{(1)},\dots,\sigma^{(k)}$), $\sigma^{(i)}\in \mathbb{R}^+$.
\end{itemize}

The concept of a Gaussian SWR model with multiple windows is illustrated in Fig.~\ref{fig:multi-window}: a target value $y_t$ is predicted by multiple windows accumulating information in ${\bm x}_{[t]}$ over different time intervals. As depicted in Fig.~\ref{fig:kernelcomb}, the combined kernel represents a linear combination of the single window kernels $\bm{\kappa}^{(i)}$ for $i=1,\dots,k$. Under the assumption that each window has a distinct centre point $\delta^{(i)}$ (otherwise, identifiability of the parameters might not be given), without loss of generality we can order the location parameters such that $\delta^{(1)}<\delta^{(2)}<\dots<\delta^{(k)}$.

\begin{figure*}[ht]
    \centering
    \begin{tikzpicture}
        \begin{groupplot}[group style={group size=1 by 2},xmin=0,ymin=0,xmax=90,height=4cm,width=\textwidth,no markers]
            \nextgroupplot[xtick={0,20,30,40,50,60,70,80,90},
                           xticklabels = {0,20,30,40,50,60,70,80,$t$},
                          ytick={0,5,10,15},
                          yticklabels={0,5,10,$y_t$~~},
                          ymax = 15]
            \addplot table [x=t, y=x, col sep=comma] {resp.csv};
            \node (a1) at (70, 10) {\textbullet};
            \draw[<->] (70, 50) -- (50, 50);
            \draw[<->] (70, 80) -- (30, 80);
            \node at (60, 60) {$\delta^{(1)}$};
            \node at (50, 90) {$\delta^{(2)}$};
            \draw[dotted] (30,80) -- (30,0);
            \draw[dotted] (50,50) -- (50,0);
            \draw[dotted] (70,80) -- (70,0);
            \nextgroupplot[xtick={0,20,30,40,50,60,70,80,90},
                          xticklabels={0,20,$s - \delta^{(2)}$,40,$s - \delta^{(1)}$,60,$s$,80, $t$},
                          ytick={0,20,40,60,80},
                          yticklabels={0,20,40,60,$x_t$~~},
                          ymax = 80]
            \draw[dashed,fill=yellow] (40, 0) rectangle ++(20,300);
            \draw[dashed,fill=yellow, opacity=0.8] (10,0) rectangle ++(40,200);
            
            \addplot table [x=t, y=x, col sep=comma] {rain.csv};
            \draw[<->] (40, 400) -- (50, 400);
            \node at (45, 500) {$r^{(1)}$};

            \draw[<->] (10, 250) -- (30, 250);
            \node at (20, 350) {$r^{(2)}$};
            
            \draw[dotted] (30,0) -- (30,1000);
            \draw[dotted] (50,0) -- (50,1000);
            \draw[dotted] (70,0) -- (70,1000);
                
            \node (b1) at (50, 200) {};
            \node (b2) at (60, 300) {};
        \end{groupplot}
        \draw[->] (b1) to [out=45,in=270] (a1);
        \draw[->] (b2) to [out=60,in=270] (a1);
    \end{tikzpicture}
    \caption{Illustration of the Sliding Windows Regression model with $k=2$ windows, which predicts $y_{70}$ based on $x_{[70]}$. The window location parameters are $\delta^{(1)} = 20$ and $\delta^{(2)} = 40$. On the time axis, the window $[40,60]$ is centred around $s-\delta^{(1)}=50$ and covers $r^{(1)} =10$ time points on each side of $\delta^{(1)}=50$, while window $[10,50]$ is centred around $s -\delta^{(2)} = 30$ and covers $r^{(2)} = 20$ time points on each side.}
    \label{fig:multi-window}
\end{figure*}

\subsubsection{Parameter interpretation}
The number of windows $k$ models the number of underlying, separable flow paths. Since single-window kernels $\bm{\kappa}$ are normalized, regression parameters (\textit{window coefficients}) $\beta^{(1)},\dots,\beta^{(k)}$ act as weights among the kernels. For practical evaluations, they may be converted into proportions representing the relative impacts of the underlying flow paths.

For a window that is essentially untruncated, the location parameter $\delta^{(i)}$ and the width $\sigma^{(i)}$ of kernel $i$ are loosely related to the mean and the standard deviation, respectively, of the time lag between precipitation and a change in streamflow. In any case $\delta^{(i)}$ is always the window mode and is thus a direct estimate of the typical time delay between rainfall events and the resulting impact on the gauged runoff for a flow path.

\begin{figure*}[ht]
    \centering
    \begin{tikzpicture}[scale=0.75]
        \draw[<->] (-7,-0.5) -- (7,-0.5); 
        \foreach \x in {1,...,12}
            \node at (-\x + 6, -0.75) {\tiny{$t$-\x}};
        \node at (6,-0.75) {\tiny{$t$}};
        \draw[dashed] (2,2) rectangle ++(6,1);
        \fill[fill=yellow] (2,2) rectangle ++(4,1);
        \node at (7.5, 2.5) {$\bm{\kappa}^{(1)}$};
        \draw[domain=2:6, smooth, samples = 100] plot (\x, {2 + exp(- (\x-5) * (\x-5) / 2)});
        \foreach \x in {2,...,6}
            \node at (\x, 2) {\textbullet};
        \draw[fill=yellow, dashed] (-6,1) rectangle ++(12,0.5);
        \node at (7.5, 1.25) {$\bm{\kappa}^{(2)}$};
        \draw[domain=-6:6, smooth, samples = 100] plot (\x, {1 + exp(-\x * \x / 8)/2)});
        \foreach \x in {-6,...,6}
            \node at (\x, 1) {\textbullet};
        \draw[fill=yellow, dashed] (-6,0) rectangle ++(12,0.5);
        \node at (8.25, 0.25) {$\beta^{(1)}\bm{\kappa}^{(1)} + \beta^{(2)}\bm{\kappa}^{(2)}$};
        \draw[domain=-6:6, smooth, samples = 100] plot ({\x}, {(exp(- \x * \x / 8) / 2 + exp(- (\x-5) * (\x-5) / 2)) / 2});
        \foreach \x in {-6,...,6}
            \node at (\x, 0) {\textbullet};
        node[right] {\tiny time};
        \node at (5,3.5) {$t-\delta^{(1)}$};
        \draw[-, dotted] (5,0) -- (5, 3);
        \node at (0,3.5) {$t-\delta^{(2)}$};
        \draw[-, dotted] (0,0) -- (0, 3);
        \draw[-, dotted] (6,0) -- (6, 3);
    \end{tikzpicture}
    \caption{Example mixture of window kernels (weights): kernel $\bm{\kappa}^{(1)}$ represents a truncated Gaussian, while kernel $\bm{\kappa}^{(2)}$ has a Gaussian shape. The combined window $\beta^{(1)}\bm{\kappa}^{(1)} + \beta^{(2)}\bm{\kappa}^{(2)}$ is a mixture of Gaussians for the SWR model.}
    \label{fig:kernelcomb}
\end{figure*}

\subsection{Information criteria}
Based on the definition and the model assumptions of the proposed Gaussian SWR model, the model errors are given by
\begin{equation*}
    \varepsilon_t = y_t - {\bm x}_{[t]} \ast \left(\sum\limits_{i=1}^{k}\beta^{(i)} \bm{\kappa}^{(i)}\right),
\end{equation*}
and follow a Gaussian white noise process. Hence, the log-likelihood $\ln L(\bm{\beta},\bm{\delta}, \bm{\sigma}; \bm{x}_T, \bm{y}_T)$ is tractable and matches that of ordinary multiple linear regression models.

To choose the number of windows, we use the Bayesian Information Criterion \citep[BIC,][]{schwarz1978bic}.
The number of model parameters in a $k$-window model aggregates one regression parameter, one window location parameter, and one window width parameter for each of the $k$ windows. Hence,  
\begin{equation}
    \text{BIC} = -2\ln L(\bm{\beta},\bm{\delta}, \bm{\sigma}; \bm{x}_T, \bm{y}_T) + \ln (\vert T \vert ) \cdot 3k,
    \label{eqn:BIC}
\end{equation}
where $\vert T\vert$ is the number of observed time points (number of samples).

\subsection{Model training \& implementation}

In a training dataset, let $\bm{y}_T$ denote the observed values of the target time series, and let $\hat{\bm{y}}_T(\bm{\beta},\bm{\delta},\bm{\sigma})$ denote the vector of predictions as a function of the model parameters; the likelihood follows from $\bm{y}_T$ and $\hat{\bm{y}}_T$. Since established approaches to fitting lagged regression models, such as \cite{hannan1967lagged} are not directly applicable due to our specific parametrization, Alg.~\ref{alg:training} outlines an iterative procedure to train the model by minimizing the negative log likelihood (or another loss function). In each iteration $k=1,\dots,k_{\max}$, one new window is added to the model and the parameters for all $k$ windows are optimized.
An information criterion (BIC for all simulations and real-data analyses) chooses the best number of windows among $1, \ldots, k_{\max}$.

\begin{algorithm*}
    \begin{algorithmic}
    \State Input:
        \State \hspace{2ex} $\bm{x}_T, \bm{y}_T$ (time series supported on a common set of times $T$) \\
        \hspace{1em} $k_{\max}$ (maximum number of windows) \\
        \hspace{1em} $n$ (population size for genetic optimization)\\
        \hspace{1em} $\bm{D}^{(0)}$ (a vector of starting values for the $\delta^{(i)}$)\\
        \hspace{1em} $\bm{S}^{(0)}$ (a vector of starting values for the $\sigma^{(i)}$)
    \State Set window counter $k\leftarrow 1$
        \State Compute $m_x$ and $m_y$, the mean of $\bm{x}_T$ and $\bm{y}_T$, respectively
        \For{$k = 1$ to $k_{max}$}
            \If{$k = 1$}
                \State Generate an $n \times 1$ column vector of starting values for each parameter in window 1 
                \State $\bm{B}^{(1)} \leftarrow (m_y / m_x, \ldots, m_y / m_x)^\top$  \Comment{(for $\beta^{(1)}$)} 
                \State $\bm{D}^{(1)} \leftarrow (\bm{D}^{(0)}, \bm{D}^{(0)}, \ldots)^\top$  \Comment{(for $\delta^{(1)}$)} 
                \State $\bm{S}^{(1)} \ \leftarrow (\bm{S}^{(0)}, \bm{S}^{(0)}, \ldots)^\top$ \Comment{(for $\sigma^{(1)}$)}  
            \Else 
                \State  Generate an $n \times 1$ vector of starting values for $\beta^{(1)}, \ldots, \beta^{(k)}$
                \For{$j = 1$ to $k$}
                    \State \hspace{1em} $\bm{B}^{(j)} \leftarrow (m_y / m_x / k, \ldots, m_y / m_x / k)$ \Comment{(for $\beta^{(j)}$)}
                \EndFor
                \State Generate an $n \times 1$ vector of starting values for $\delta^{(k)}$ (note $\bm{D}^{(1)}, \ldots, \bm{D}^{(k-1)}$ are unchanged)
                \State $\bm{D}^{(k)} \leftarrow \text{permute}(\bm{D}^{(1)})$ 
                \State Generate an $n \times 1$ vector of starting values for $\sigma^{(k)}$ (note $\bm{S}^{(1)}, \ldots, \bm{S}^{(k-1)}$ are unchanged)
                \State $\bm{S}^{(k)} \leftarrow \text{permute}(\bm{S}^{(1)})$ 
            \EndIf
            \State Generate an $n \times (3k)$ population of starting values with columns for $\beta^{(1)}, \ldots, \beta^{(k)}, \delta^{(1)}, \ldots, \delta^{(k)}, \sigma^{(1)}, \ldots, \sigma^{(k)}$
            $$P = \left(\begin{array}{c|c|c|c|c|c|c|c|c} 
            \bm{B}^{(1)} & \cdots & \bm{B}^{(k)} & \bm{D}^{(1)} & \cdots & \bm{D}^{(k)} & \bm{S}^{(1)} & \cdots & \bm{S}^{(k)}
            \end{array} \right)$$ 
            \State Minimize the negative log likelihood for $k$ windows via the GENOUD algorithm with starting values in $P$   
            $$(\hat{\bm{\beta}}^{(k)}, \hat{\bm{\delta}}^{(k)}, \hat{\bm{\sigma}}^{(k)}) \leftarrow \underset{\bm{\beta}, \bm{\delta}, \bm{\sigma}}{\arg \min}
            -\ln L(\bm{\beta},\bm{\delta}, \bm{\sigma}; \bm{x}_T, \bm{y}_T).$$
            \State Compute $\text{BIC}_k$ for $k$ windows from (\ref{eqn:BIC})
        \EndFor
        \State return $\hat{k}$ and $\hat{\bm{\beta}}^{(\hat{k})}, \hat{\bm{\delta}}^{(\hat{k})}, \hat{\bm{\sigma}}^{(\hat{k})}$, such that information criterion $\text{BIC}_k$ is optimized 
        $$\hat{k} = \underset{k \in \{1, ..., k_{max}\}}{\arg \min}\text{BIC}_k$$ 
    \end{algorithmic}
    \caption{Train a Gaussian SWR model, choosing the number of windows}
    \label{alg:training}
\end{algorithm*}

We use the GENOUD algorithm \citep{mebane2011genoud} in the {\tt R} package {\tt rgenoud} \citep{rgenoud} to train a model with a fixed number of windows.
GENOUD optimizes a population of candidate solutions over a number of generations via operations analogous to genetic evolution. It also fine tunes the best candidate found at each generation with a quasi-Newton method. 
Trial and error comparisons with a more standard constrained optimizer \citep[BOBYQA, see][]{powell2009bobyqa} demonstrated a clear advantage in the optimal log likelihood found, which is important if BIC is to reliably choose the number of windows. 
As a stopping criterion, the absolute tolerance is set to $10^{-3}$,
which is a trivial change in the log likelihood.

In the GENOUD algorithm, we constrain the parameters such that all $\delta^{(i)} > 0$ and all $\sigma^{(i)} > 1/6$, which allows for truncated as well as non-truncated kernels. In the boundary case of $\delta=0$, a maximum level of truncation is reached for the Gaussian kernel. The constraint $\sigma^{(i)}  > 1/6$, avoids collapse of the window.  Moreover, the minimum window width is hence 1, i.e., $\pm 3 \times 1/6$, which is commensurate with the limitation of discrete data at unit time steps.

GENOUD requires a population of $n$ starting values for $\beta$, $\delta$, and $\sigma$. Experimental observations showed that this initial population is important. It needs to provide coverage of reasonable values in the parameter space but not be so overly wide as to make the optimization inefficient. Note that the initial ranges can be exceeded later during the GENOUD iterations. The population formed in Alg.~\ref{alg:training} has $n$ members in the rows of a matrix ($n = 100$ for all results reported) and $3k$ columns for initial values of $\beta^{(1)}, \ldots, \beta^{(k)}, \delta^{(1)}, \ldots, \delta^{(k)}, \sigma^{(1)}, \ldots, \sigma^{(k)}$, respectively. For $k = 1$, $\beta^{(1)}$ is always set to $m_y / m_x$, where $m_x$ and $m_y$ are the sample means of $\bm{x}_T$ and $\bm{y}_T$, respectively, to respect the scaling of the data when a single kernel has total unit weight. The $\delta^{(1)}$ starting values repeat a vector $\bm{D}^{(0)}$; we use $\bm{D}^{(0)} = (1, 2, \ldots, 25)^T$, which will be repeated four times to fill the population of size $n = 100$.  This range is reasonable for the hydrology application but would be changed in other contexts. 
Similarly, the $\sigma^{(1)}$ starting values repeat $\bm{S}^{(0)} = (1, 5, 10, 20)^T$.  When $k > 1$ and window $k$ is added, $\beta^{(1)}, \ldots, \beta^{(k)}$ are all reduced in magnitude to $m_y / m_x / k$ to account for $k$ kernels.  The starting values for $\delta^{(1)}$ and $\sigma^{(1)}$ are as for the $k = 1$ iteration; the $\delta^{(k)}$ and $\sigma^{(k)}$ for the further windows are given by a random permutation of the elements in $\bm{D}^{(0)}$ and $\bm{S}^{(0)}$, respectively. This procedure ensures the starting values for a multi-window iteration have sufficient coverage of the parameter space, while having distinct initial values for each of the $k$ windows.

If the number of windows is not known a priori, an upper bound $k_{\max}$ must be defined, and the hyperparameter $\hat{k}$ will be selected by choosing the best among all $k_{\max}$ iterations based on BIC.

Note that the suggested model initializations are specific to the hydrological application tackled in this work. A different choice of the hyperparameters may be required if the model was transferred to another domain. The Gaussian SWR model is implemented in R version 4.3.0 \citep{r2022}. The implementation is publicly available on GitHub\footnote{\url{https://www.github.com/sschrunner/SlidingWindowReg}}.

\subsection{Autocorrelated residuals}

We expect autocorrelation to be present in the residuals of the Gaussian SWR model, since both input and output time series may have other effects or measurement errors correlated with time. This aspect violates the requirements of least squares likelihood and may distract evaluation metrics such as mean squared error. Thus, we aim to resolve such issues by adapting the estimation procedure proposed by \citet{cochrane1949}.

Suppose the errors $\varepsilon_t$ in the model~(\ref{eq:multiwindow}) follow an autoregressive process of lag 1, i.e., AR(1), so that
\begin{equation}
    \varepsilon_t = \varphi \varepsilon_{t-1} + \eta_t, \label{eq:ar1}
\end{equation}
where $\eta_t$ is Gaussian white noise (instead of $\varepsilon_t$ in Eq.~(\ref{eq:multiwindow})), and $\varphi\in (-1,1)$. From Eq.~(\ref{eq:multiwindow}) and (\ref{eq:ar1}), it follows that 
\begin{align*} y_t &= \sum_{i=1}^k \beta^{(i)} \cdot \left({\bm x}_{[t]} \ast \bm{\kappa}^{(i)}\right) + \varphi \varepsilon_{t-1} + \eta_t, \text{and}\\
y_{t-1} &= \sum_{i=1}^k \beta^{(i)} \cdot \left({\bm x}_{[t-1]} \ast \bm{\kappa}^{(i)}\right) +  \varepsilon_{t-1}
\end{align*}
In order to convert this setup into a model with Gaussian white noise errors, we investigate the expression
\begin{align*}
    y_t - \varphi y_{t-1} &= \sum_{i=1}^k \beta^{(i)} \cdot \left(\bm x_{[t]} \ast \bm\kappa^{(i)} - \varphi \bm x_{[t-1]} \ast \bm\kappa^{(i)} \right) + \eta_t \\
    &=\sum_{i=1}^k \beta^{(i)} \cdot \left(\sum\limits_{s=1}^{t}\left(x_s - \varphi x_{s-1}\right) \kappa^{(i)}_{t-s} \right) + \eta_t.
\end{align*}
The term $x_0 \cdot \kappa^{(i)}_{t}$ originating from index $s=0$ is ignored since $\bm\kappa^{(i)}$ is padded with $0$ at the end (when being adjusted to the length of $\bm x_{[t]}$). Hence, given $\varphi$, the model can be written as an ordinary least squares model with uncorrelated errors $\eta_t$: 
$$ \tilde{y}_t = \sum_{i=1}^k \beta^{(i)} \cdot \left(\tilde{\bm{x}}_{[t]} \ast \bm{\kappa}^{(i)} \right) + \eta_t,$$
where the transformation $\tilde{z}_t = z_t - \varphi z_{t-1}$ can be applied a priori as part of the preprocessing procedure on both variables $x_t$ and $y_t$.

For the Gaussian SWR model, we adapt the Cochrane-Orcutt procedure to estimate $\varphi$ and the optimal model parameters:
\begin{itemize}
    \item estimate model parameters $\bm{\beta}, \bm{\delta},\bm{\sigma}$ from the original model using $\bm{x}_{[t]}$ and $\bm{y}_{[t]}$ (i.e., set $\varphi = 0$),
    \item estimate $\varphi$ from the model residuals,
    \item apply the transformation $\tilde{z}_t = z_t - \varphi z_{t-1}$ to both $\bm{x}_T$ and $\bm{y}_T$ and obtain final estimates of the model parameters $\bm{\beta}, \bm{\delta},\bm{\sigma}$.
\end{itemize}
The same concept also holds for more general AR($m$)-processes with $m>1$. In general, the transformation leads to a loss of $m$ time points (samples) in the training dataset.

\subsection{Uncertainty quantification}

Minimizing the negative log-likelihood with respect to $(\bm{\beta}, \bm{\delta}, \bm{\sigma})$  gives maximum-likelihood estimates of the model parameters. Large-sample theory implies the errors in the estimators have an approximate multivariate Gaussian distribution with mean $\bm{0}$ (asymptotically unbiased) and an approximate covariance matrix $\mathcal{I}^{-1}$, where 
$$\mathcal{I}(\bm{\beta}, \bm{\delta}, \bm{\sigma}) = \frac{\partial^2}{\partial (\bm{\beta}, \bm{\delta}, \bm{\sigma})^2} \left( -\ln L(\bm{\beta}, \bm{\delta}, \bm{\sigma};\bm{x}_T, \bm{y}_T) \right)$$
denotes the observed information, i.e., the negative second derivative of the log-likelihood evaluated at the MLE parameter values. 
The Hessian can be obtained numerically via the R package \texttt{numDeriv}~\citep{numDeriv2019}. The diagonal elements of $\mathcal{I}^{-1}$ lead to approximate standard errors, which provide the uncertainty quantification reported below for the two watersheds studied. These watersheds have very long time series to support the asymptotic argument.

\section{Simulation Study}
\label{sec:simulationstudy}

In a simulation study, the proposed Gaussian SWR model will be validated in controlled scenarios based on real-world input data and simulated targets.

\subsection{Experimental Setup}

Precipitation data $\bm{x}_T$ is used from the Koksilah River watershed which is located in Cowichan, British Columbia, Canada, while an artificial target variable $\bm{y}_T$ is sampled based on Eq.~(\ref{eq:multiwindow}). Both time series are acquired on a daily basis. The model parameters $\bm{\beta}$, $\bm{\delta}$ and $\bm{\sigma}$ are randomly generated for this purpose, along with Gaussian white noise errors $\varepsilon_t\sim N(0,\rho^2)$. The distributions of the model parameters and level of the error variance $\rho^2$ are described further below.

In all of the following setups, the model parameters of the Gaussian SWR model were estimated on a training set comprising a time series of the first 29 hydrological years, resulting in 10,593 data points. The test set consisted of the remaining 10 time series of hydrological years or 3,652 data points, which corresponds to an approximate 75\%/25\% split of the dataset.

The primary factors steering the difficulty associated with a setup in the simulation study are:
\begin{enumerate}
    \item the (ground truth) number of windows $k^{\text{gt}}$ --- the dimensionality and complexity of parameter estimation increases with the number of windows,
    \item the pairwise overlap between (ground truth) windows on the time axis --- a higher overlap leads to a decrease in the separability of windows, and
    \item the level of measurement noise $\rho^2$ applied, given by the noise rate (relative to the explained variance of the model, as highlighted below).
\end{enumerate}
While $k^{\text{gt}}$ and $\rho$ are systematically varied, the overlap of windows is a result from the random sampling of ground truth model parameters.

We employ 15 distinct setups (five 1-, 2-, and 3-window setups, respectively). A higher number of windows is not common for hydrological models due to the limited number of possible flow paths. Each setup uses a set of parameters independently sampled from $\delta^{(i)}\sim U(0,20)$, and $\sigma^{(i)},\beta^i\sim U(1,5)$ for any $i\in \{1,\dots,k\}$, where $U(a,b)$ denotes a uniform distribution on the interval $[a,b]$. Each 1-, 2- and 3-window setup is repeated at 5 distinct levels of measurement noise $\alpha\in \{0.05, 0.25,0.5,0.75,0.95\}$ to control the error relative to the signal, for a total of 75 simulation experiments.

Given a noise level $\alpha$, the measurement noise is added as follows. After evaluating the deterministic terms in Eq.~(\ref{eq:multiwindow}) with the given model parameters, denoted as $\tilde{\bm{y}}_T$, we compute the associated sample variance $\text{Var}(\tilde{\bm{y}}_T)$. Gaussian white noise with standard deviation $\rho = \alpha \cdot \sqrt{\text{Var}(\tilde{\bm{y}}_T)}$ is simulated and added to construct $\bm{y}_T = \tilde{\bm{y}}_T + \bm{\varepsilon}_T$.

The resulting sample variance of $\bm{y}_T$ can be expressed as $\text{Var}(\bm{y}_T) = (1+\alpha^2)\text{Var}(\tilde{\bm{y}}_T)$. Thus, $\alpha^2$ represents the relative noise level with respect to the explained variance. This induces an upper bound on the $R^2$ score~\citep{kuhn2019applied} achievable by any predictor given $\bm{x}_T$ and $\bm{y}_T$: a perfect fit giving $\hat{\bm{y}}_t = \tilde{\bm{y}}_T$ would achieve an $R^2$ score of 
\begin{align}
    R^2\left(\bm{y}_T,\hat{\bm{y}}_T\right) &=  1 - \frac{\Vert \bm{y}_T-\hat{\bm{y}}_T\Vert_2^2}{\Vert \bm{y}_T-\bar{y}\bm{1}\Vert_2^2} \nonumber\\
    &= 1 - \frac{\text{Var}(\bm{\varepsilon}_T)}{\text{Var}(\bm{y}_T)} \nonumber\\
    &= 1 - \frac{\alpha^2\text{Var}(\tilde{\bm{y}}_T)}{(1+\alpha^2)\text{Var}(\tilde{\bm{y}}_T)} \nonumber\\
    &= \frac{1}{1+\alpha^2}, \label{eq:r2}
\end{align}
where $\bar{y}$ denotes the mean of $\bm{y}_T$ across $T$, and $\bm{1}$ denotes a vector of ones. All noise levels $\alpha$ and associated $R^2$ scores are shown in Table~\ref{tab:setup}. 

\begin{table}[ht]
    \centering
    \caption{Noise levels and associated maximum $R^2$ scores (upper bound for predictive model performance).}
    \label{tab:setup}
    \begin{tabular}{lll}
           \toprule
            & noise level $\alpha$ & maximum $R^2$ \\
           \midrule
           1 & 0.05 & 0.998 \\
           2 & 0.25 & 0.941 \\
           3 & 0.5 & 0.8 \\
           4 & 0.75 & 0.64 \\
           5 & 0.95 & 0.526 \\
            \bottomrule
    \end{tabular}
\end{table}

\subsection{Evaluation Metrics}
In order to evaluate the Gaussian SWR model, we deploy evaluation metrics to assess both the accuracy of parameter estimation and the predictive performance. To facilitate the assessment of the estimated parameters, we introduce the notion of \textit{overlap} between two weight vectors with equal dimensions, $\bm{w}^{(1)},\bm{w}^{(2)}\in\mathbb{R}^t$ as follows:
$$ O(\bm{w}^{(1)},\bm{w}^{(2)}) = \sum\limits_{s=1}^{t} \min\{ w_s^{(1)}, w_s^{(2)}\}. $$
Hence, in a functional setting, the overlap would describe the area covered by the minimum of both kernel functions. Since both weight vectors are positive and sum to 1 due to the normalization condition, the overlap will be in $[0,1]$ and reaches $1$ only if both weight vectors are exactly equal. For a multiple-window setting, a combined kernel takes account of the kernel weights but is then normalized for ease of comparison when computing the overlap.

We group the evaluation criteria applied to our model results into two categories:
\begin{itemize}
    \item \textbf{Kernel overlap}: When combining all estimated windows to a joint weight vector $$\sum\limits_{i=1}^k \beta^i \bm{\kappa}^{(i)},$$ i.e., the mixture of all kernels, the overlap between the full predicted and the ground truth kernels is assessed (including the weighting of windows by regression parameters $\bm{\beta}$). In particular, we compute the overlap $O(\hat{\bm{\theta}},\bm{\theta})$, where $\hat{\bm{\theta}} = (\hat{\bm{\beta}}, \hat{\bm{\delta}}, \hat{\bm{\sigma}})$, and $\bm{\theta} = (\bm{\beta}, \bm{\delta}, \bm{\sigma})$, respectively. Thus, the combined kernel overlap evaluates the quality of overall parameter estimation, including the regression parameters $\bm{\beta}$.
    \item \textbf{Predictive performance}: Conventional regression metrics are used to evaluate the predictive performance of the model on the test set \citep{kuhn2019applied}. These include the root mean squared error (RMSE), as well as the coefficient of determination ($R^2$) on the test set. The latter is also known as Nash-Sutcliffe Efficiency (NSE) in hydrology~\citep{nash1970nsc} and is introduced in Eq.~(\ref{eq:r2}). In addition, the Kling-Gupta efficiency (KGE) is a performance metric used in hydrology which, unlike the NSE, independently encourages predictions to match the variability of observations, thereby removing the tendency of NSE to underestimate high flows and overestimate low flows \citep{gupta2009kge}. It uses the following formulation:
    \noindent \begin{equation*} 1 - \sqrt{(r-1)^2 + \left(\frac{\sigma_{\hat{\bm{y}}_T}}{\sigma_{\bm{y}_T}} - 1\right)^2 + \left(\frac{\mu_{\hat{\bm{y}}_T}}{\mu_{\bm{y}_T}} - 1\right)^2},\end{equation*}\noindent
    where $r$ denotes the Pearson correlation coefficient between $\bm{y}_T$ and $\hat{\bm{y}}_T$, $\mu$ represents the respective means, and $\sigma$ represents the respective standard deviations.
    All performance metrics are implemented using the R package {\tt hydroGOF} \citep{bigiarini2020hydroGOF}.
\end{itemize} 

\subsection{Results for uncorrelated errors}

First, we visualize the predicted kernels obtained from training the Gaussian SWR model in each setup specified above. In the experiment, the algorithm selects the number of windows $k$ based on the best BIC, subject to a maximum of $k_{\max} = 3$ windows. This upper limit originates from our application, where typically a maximum number of up to 3 distinct flow paths (overland, sub-surface, ground water flow) is assumed.

\begin{figure}[ht]
    \centering
    \includegraphics[width = 0.65\textwidth]{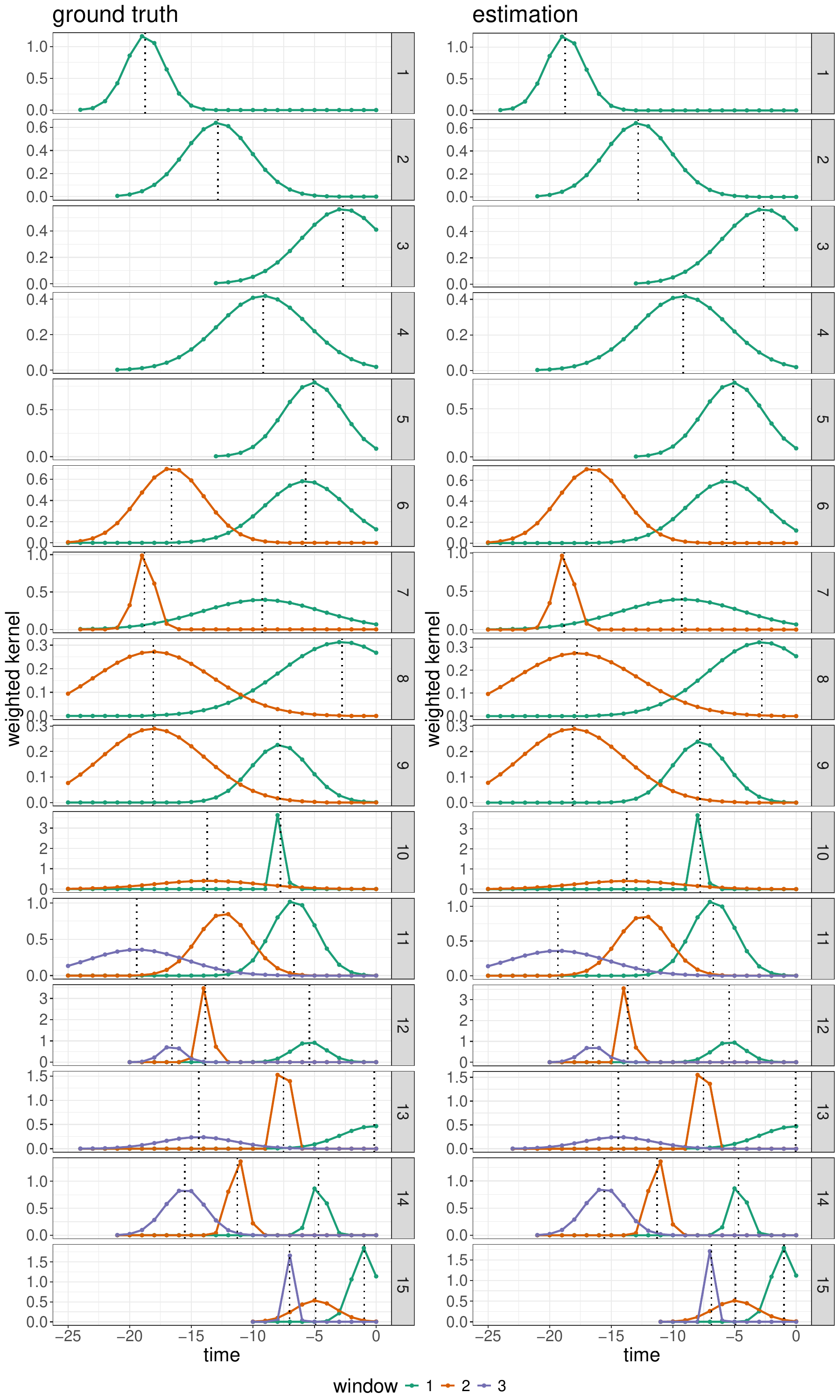}
    \caption{Weights at discrete time points of the ground truth versus predicted kernels
        (noise level $\alpha = 0.5$). The number of windows was selected as a model hyperparameter.}
    \label{fig:kernel_comparison}
\end{figure}

Fig.~\ref{fig:kernel_comparison} compares the weights at discrete time points of the ground truth and estimated kernels for the middle noise level $\alpha = 0.5$ in each setup. Dotted vertical lines indicate the true or estimated window centres $\delta^{(i)}$ along the time axis. In all 1-, 2-, and 3-window setups, window positions and sizes were accurately predicted. Moreover, dominant peaks and general shapes are reconstructed in all scenarios. Since we allow non-integer numbers for delta, the discretized weights are not always symmetric w.r.t. the measured time points. Overall, the combined kernel matches with the ground truth very accurately, and the general separation of the windows is acceptable.

Beyond visual comparison, Table \ref{tab:mod_param_estim} provides a summary of window overlap accuracy based on predicted and true parameters. Overall, all setups achieve excellent overlaps with the ground truth kernels. Thus, the reconstruction of the overall weighted kernel is at a high level across all simulated setups.

\begin{table}[ht]
    \centering
    \caption{Parameter estimation accuracy for 1-, 2- and 3-window setups across five noise levels.  An overlap of 1 indicates perfect agreement between an estimated window and the ground truth.}
    \label{tab:mod_param_estim}
    \begin{tabular}{l|lllll}
\toprule
& \multicolumn{5}{c}{noise level $\alpha$} \\
\cline{2-6}
           setup no. & 0.05 & 0.25 & 0.5 & 0.75 & 0.95 \\
\midrule
    \multicolumn{2}{c}{1-window setups} \\
    \midrule
     1 & 1.00 & 1.00 & 1.00 & 1.00 & 0.99 \\ 
     2 & 1.00 & 1.00 & 1.00 & 1.00 & 0.99 \\ 
     3 & 1.00 & 1.00 & 1.00 & 1.00 & 1.00 \\ 
     4 & 1.00 & 1.00 & 1.00 & 0.99 & 1.00 \\ 
     5 & 1.00 & 1.00 & 0.99 & 1.00 & 0.99 \\ 
    \midrule
     mean & 1.00 & 1.00 & 1.00 & 1.00 & 0.99 \\ 
    \midrule
    \multicolumn{2}{c}{2-window setups} \\
    \midrule
      6 & 1.00 & 1.00 & 0.99 & 0.99 & 0.99 \\  
      7 & 1.00 & 1.00 & 0.99 & 0.99 & 0.99 \\
      8 & 1.00 & 1.00 & 0.99 & 0.99 & 0.99 \\ 
      9 & 1.00 & 1.00 & 0.99 & 0.99 & 0.98 \\
      10 & 1.00 & 1.00 & 1.00 & 0.99 & 0.98 \\ 
    \midrule
     mean & 1.00 & 1.00 & 0.99 & 0.99 & 0.99 \\ 
    \midrule
    \multicolumn{2}{c}{3-window setups} \\
    \midrule
      11 & 1.00 & 0.99 & 0.99 & 0.98 & 0.98 \\ 
      12 & 0.99 & 0.99 & 0.98 & 0.98 & 0.97 \\  
      13 & 1.00 & 1.00 & 0.99 & 1.00 & 0.97 \\
      14 & 1.00 & 0.98 & 0.99 & 0.98 & 0.98 \\ 
      15 & 1.00 & 0.99 & 0.98 & 0.97 & 0.96 \\ 
    \midrule
     mean & 1.00 & 0.99 & 0.99 & 0.98 & 0.97 \\ 
    \bottomrule
\end{tabular}

\end{table}

\begin{figure}[ht]
\centering
    \includegraphics[width=0.4\textwidth]{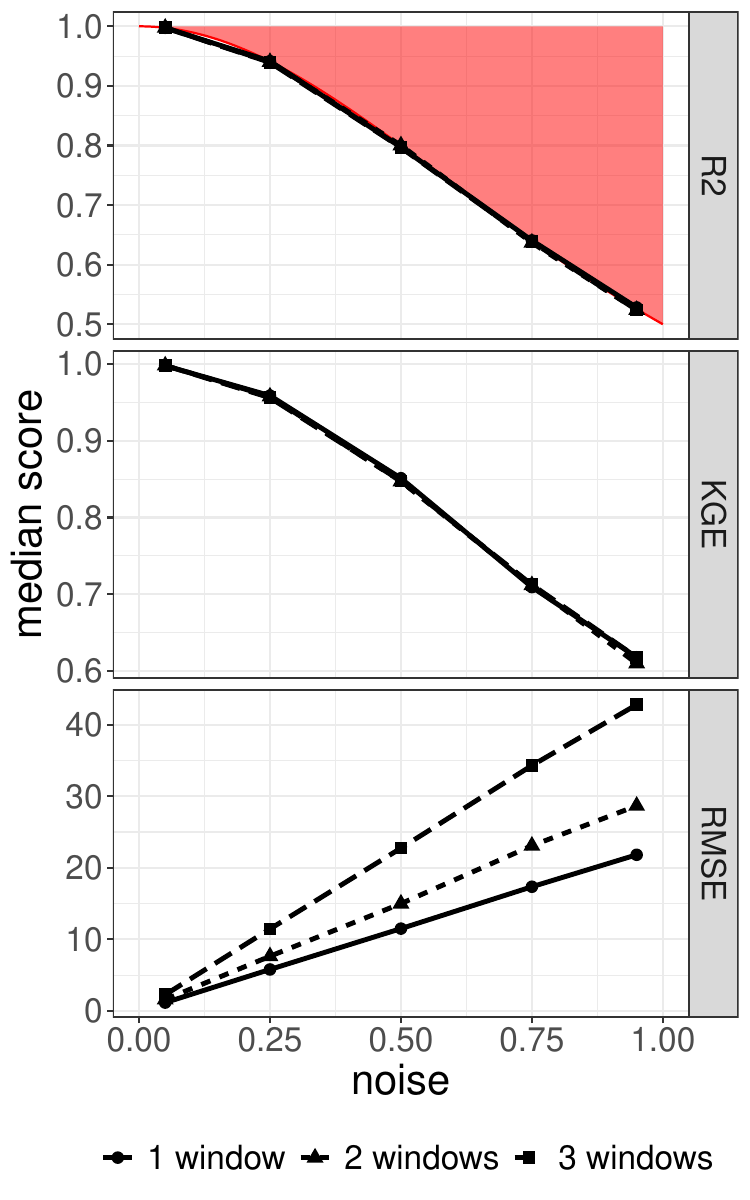}
    \caption{Predictive performance summaries for 1-, 2- and 3-window setups by noise level $\alpha$. All results are averaged over the 5 distinct simulated parameter setups.}
    \label{fig:sim_predictive_metrics}
\end{figure}

Having observed the high accuracy of parameter identification in the simulation experiment, we further investigate the predictive performance of the models with respect to the target variable on the test set. Figure~\ref{fig:sim_predictive_metrics} illustrates the corresponding $R^2$, KGE, and RMSE scores obtained by the estimated Gaussian SWR models. In agreement with the observations made on the quality of parameter estimation, the predictive performance remains at a high level and is mainly affected by the noise levels. For the $R^2$ score, the upper bounds in Table \ref{tab:setup} are indicated by a red region, which cannot be reached at the given noise level. The achieved average $R^2$ values are close to the upper bounds across all model setups,  and hence the estimated models obtain almost optimal prediction accuracy on the test set.

To guarantee interpretability of the model parameters, accurate estimation of the true number of windows $k^{\text{gt}}$ is crucial. For $k^{\text{gt}}$ equal to 1, 2, or 3, all 75 estimated models select the hyperparameter $k$ correctly. 

\subsection{Results for autocorrelated errors}

To account for a more general error structure, we repeat the above experiment but with response data simulated with autocorrelated AR(1) errors $\varepsilon_t$. We make use of the R function \texttt{arima.sim} from the \texttt{forecast} package ~\citep{hyndman2021arima} to simulate AR(1) noise  with a given standard deviation $\rho$.  Analogous to the case of uncorrelated errors, $\rho$ is based on the sample variance $\text{Var}(\tilde{y}_T)$ and the noise levels $\alpha$ from Table~\ref{tab:setup}. Internally, the function constructs i.i.d. errors $\eta_t$ with the given standard deviation $\rho$, which are then transformed to form an AR(1) process. The true autoregressive parameter $\varphi$ is set to 0.5 throughout for the purpose of demonstrating the concept.

In practice, the autocorrelation has to be assessed, and the autoregressive parameter $\varphi$ needs to be estimated when we extend the Gaussian SWR model. The Durbin-Watson test for autocorrelated model errors is performed to assess the residuals before and after the Cochrane-Orcutt data transformation. As expected, model fits with no Cochrane-Orcutt transformation deliver p-values below $0.01$ (rejecting the null hypothesis of no autocorrelation) for all 75 setups.  The average absolute estimation error of the parameter $\varphi$ is less than $0.006$ across all setups, suggesting highly accurate estimation. After applying the Cochrane-Orcutt procedure, 74 out of the 75 setups lead to residuals with p-values above $0.05$, which provides no evidence for violation of the assumption of uncorrelated residuals in the model fit to transformed data. The remaining setup, with p-value of 0.032, has only small residual autocorrelation of $0.02$. 

With autocorrelated errors, the upper bounds for the $R^2$ scores achievable by the model must be modified. 
From~(\ref{eq:ar1}) we have $\text{Var}(\varepsilon_t)=\text{Var}(\varphi \varepsilon_{t-1} + \eta_t) 
= \varphi^2 \text{Var}(\varepsilon_{t-1}) + \text{Var}(\eta_t)$ as $\eta_t$ is uncorrelated with $\varepsilon_{t-1}$ by construction.
Iterating this result, and noting that
$\text{Var}(\varepsilon_1) = \text{Var}(\eta_1)$ and $\text{Var}(\eta_t)=\alpha^2 \text{Var}(\tilde{\bm{y}}_T)$, gives 
\begin{align*}
    \text{Var}(\varepsilon_t) &= \varphi^2\text{Var}(\varepsilon_{t-1}) + \text{Var}(\eta_{t}) \\
    &= \varphi^4\text{Var}(\varepsilon_{t-2}) + \varphi^2\text{Var}(\eta_{t-1}) + \text{Var}(\eta_{t}) \\
    &= \sum\limits_{i=0}^{t-1} \varphi^{2i} \text{Var}(\eta_{t-i}) \\
    &= \underbrace{\frac{1-\varphi^{2(t-1)}}{1-\varphi^2}}_{\xi} \alpha^2 \text{Var}(\tilde{\bm{y}}_T).
\end{align*}
It holds that $\xi\approx \frac{1}{1-\varphi^2}$ for longer time series.
As a result, the maximum achievable $R^2$ score (when $\hat{\bm{y}}_T = \tilde{\bm{y}}_T$) under the AR(1) model errors is given by
\begin{align*} 
    R^2(\bm{y}_T,\hat{\bm{y}}_T) &= 1 - \frac{\xi \alpha^2 \text{Var}(\tilde{\bm{y}}_T)}{(1 + \xi \alpha^2)\text{Var}(\tilde{\bm{y}}_T)} \\ 
    &= \frac{1}{1 + \xi \alpha^2}.
\end{align*}

\begin{figure}[ht]
\centering
    \includegraphics[width=0.4\textwidth]{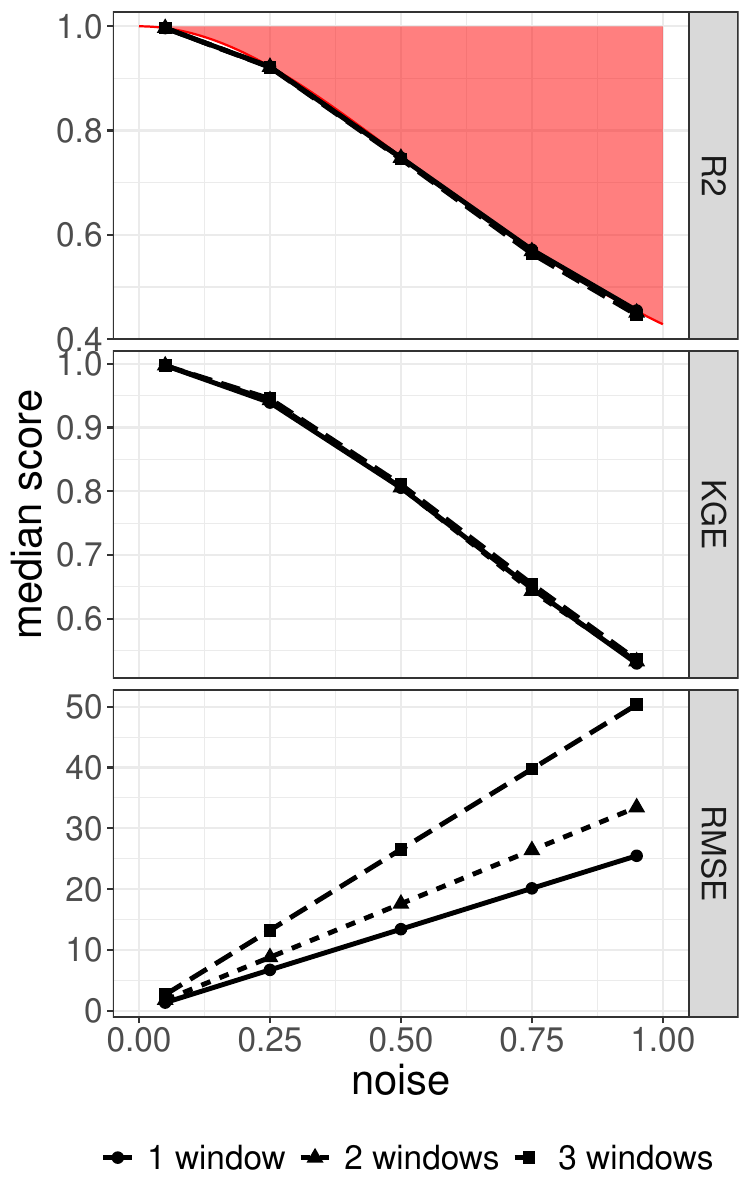}
    \caption{Predictive performance summaries for 1-, 2- and 3-window setups by noise level $\alpha$ in the autocorrelated setup}
    \label{fig:sim_predictive_metrics1}
\end{figure}

Similar to the simulation with uncorrelated residuals, the estimated kernels closely align with the ground truth, demonstrating accurate reconstruction of the overall weighted kernel across all simulated setups. For the given noise levels $\alpha$, the maximum $R^2$ scores for the autocorrelated simulations are presented in Table~\ref{tab:setup1}. Results obtained for the experimental setups are shown in Fig.~\ref{fig:sim_predictive_metrics1}. As with the uncorrelated model setups, the $R^2$ scores almost reach the theoretical upper bound.

\begin{table}[ht]
    \centering
    \caption{Noise levels and associated maximum $R^2$ scores (upper bound for predictive model performance) in the autocorrelated setups.}
    \label{tab:setup1}
    \begin{tabular}{lll}
           \toprule
            & noise level $\alpha$ & maximum $R^2$ \\
           \midrule
           1 & 0.05 & 0.997 \\
           2 & 0.25 & 0.923 \\
           3 & 0.5 & 0.75 \\
           4 & 0.75 & 0.571 \\
           5 & 0.95 & 0.454 \\
            \bottomrule
    \end{tabular}
\end{table}

Overall, the hyperparameter $k$ representing the number of selected windows, determined via BIC matches the ground truth in all 1-window setups, overestimates by 1 for 3 out of 25 of the 2-window setups, and underestimates by 1 for one of the 25 3-window simulations.

\section{Real-World Data Experiments}
\label{sec:experiments}
In the second part of our experiments, we evaluate the performance of the Gaussian SWR model on two real-world examples. Both watersheds are located on the west coast of North America, with the first catchment representing the the Koksilah River which is located in Cowichan, British Columbia, Canada. The second catchment represents Big Sur River in central California, United States. The available dataset consists of 39 hydrological years for both watersheds. The Koksilah River has been investigated in previous works~\citep{janssen2023learning}. Both datasets are available on the aforementioned GitHub repository (entitled \texttt{sampleWatershed} and \texttt{sampleWatershed2}, respectively).

\subsection{Experimental Setup}
Analogous to the simulation study, we use a training subset containing years 1 to 29 of the time series, followed by a test set covering 10 years. The maximum number of windows is set to $k_{\max}=4$.

After initial training, the model residuals are evaluated for autocorrelation using the Durbin-Watson test. If autocorrelation is detected, the Cochrane-Orcutt procedure is applied to correct for AR(1) errors. If the Durbin-Watson test finds no evidence of autocorrelation in the the updated model the process stops;  otherwise AR(2) errors are considered, etc.

\subsection{Results}

The Durbin-Watson test on the residuals from the model with untransformed data gives a p-value $<0.01$ in both cases, i.e. strong evidence against an independent-error model. In the case of watershed 1, an AR($2$) process with parameters $\varphi_1=0.46$ and $\varphi_2 = 0.12$ is indicated to model the autocorrelation, resulting in a Durbin-Watson p-value of $0.20$ after applying the Cochrane-Orcutt procedure. For watershed 2, an AR($1$) process with parameter $\varphi = 0.84$ appears to be sufficient to remove the autocorrelation, resulting in a p-value of $0.39$ after data transformation.

\begin{figure*}[ht]
    \centering
    \begin{subfigure}{0.7\textwidth}
        \centering
        \includegraphics[width = \textwidth]{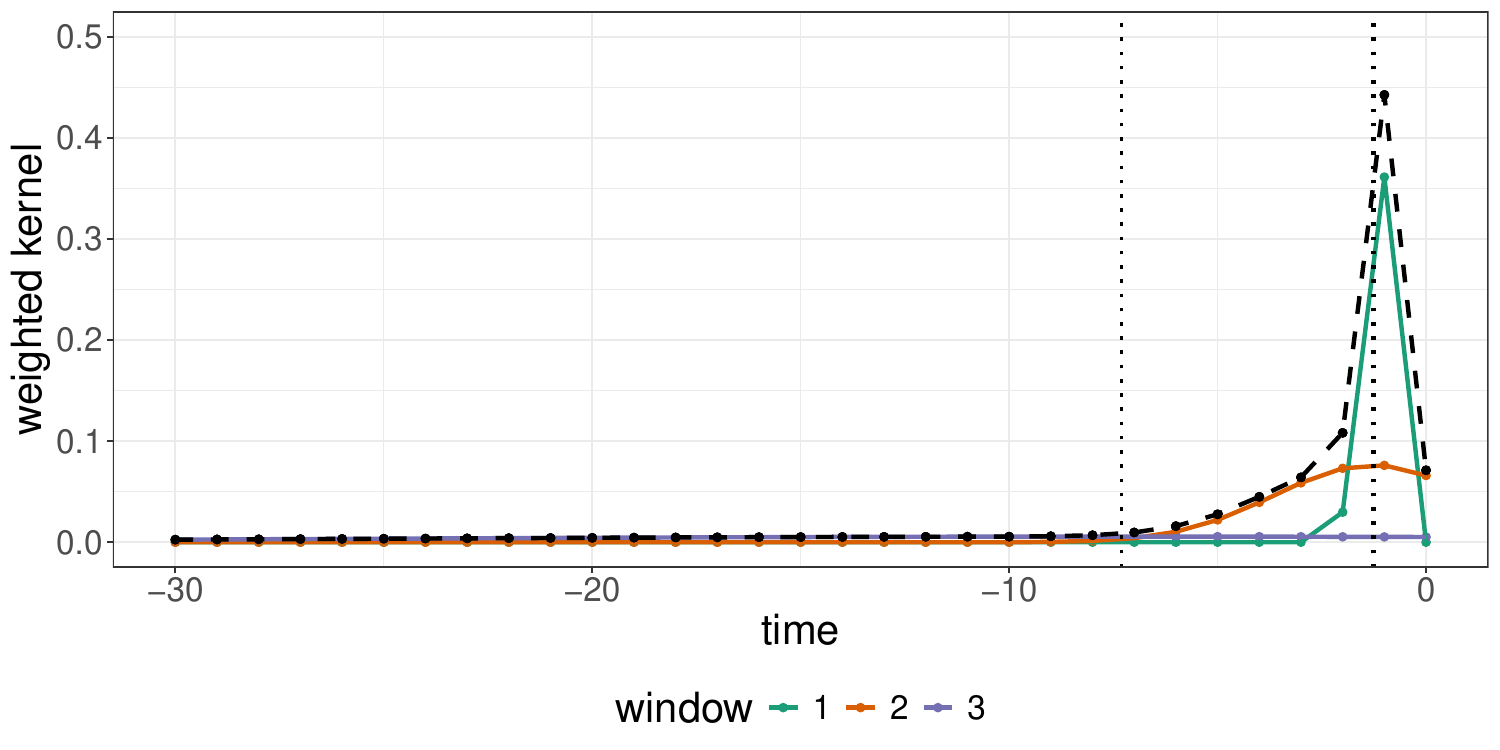}
        \caption{Koksilah River (Watershed 1)}
        \label{fig:real_world_kernels1}
    \end{subfigure}
    \begin{subfigure}{0.7\textwidth}
        \centering
        \includegraphics[width = \textwidth]{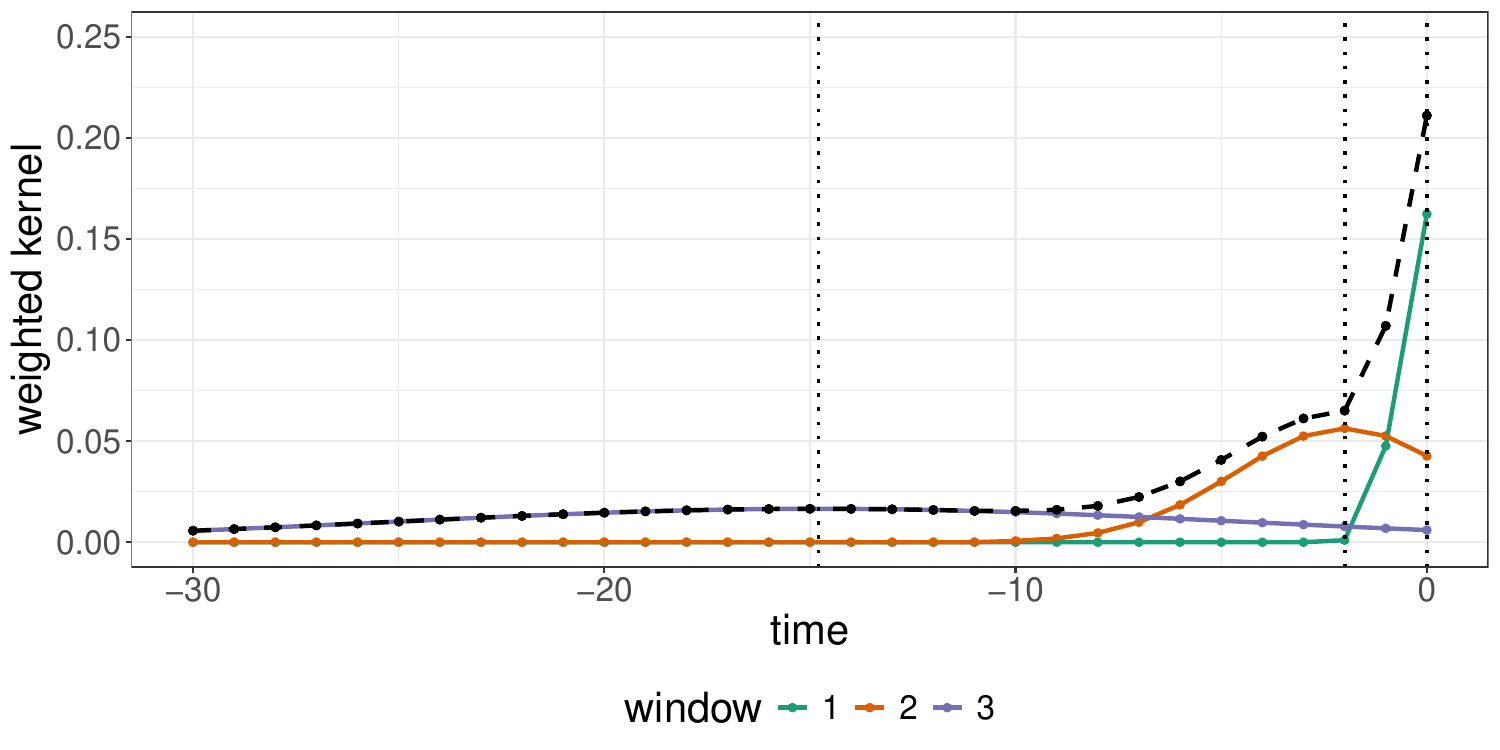}
        \caption{Big Sur River (Watershed 2)}
        \label{fig:real_world_kernels2}
    \end{subfigure}
    \caption{Estimated weights at discrete time points of scaled window kernels in real-world data experiments. The combined kernel (dashed line) for watershed 1 resembles a truncated Student-$t$ distribution with stronger tails than Gaussian probability density functions. The combined kernel (dashed line)  for watershed 2 is multimodal.}
    \label{fig:real_world_kernels}
\end{figure*}

The estimated kernels for each watershed are illustrated in 
Fig.~\ref{fig:real_world_kernels}, and their parameters are shown along with uncertainties in Table~\ref{tab:realworld_params}. Predictive performance scores ($R^2$, $KGE$, and RMSE) on the test set are shown in Table~\ref{tab:realworld_perf}. Overall, the Gaussian SWR model achieves reasonably accurate predictions for both watersheds, e.g., $R^2$ scores of approximately 0.71 and 0.6, respectively, on the untransformed streamflow scale. 

\begin{table*}[ht]
    \centering
    \caption{Parameter estimates after Cochrane-Orcutt transformation for real-world watersheds, with approximate standard errors.}
    \label{tab:realworld_params}
    \begin{tabular}{ll|rrr}
           \toprule
            watershed no & window no & $\beta^{(i)}$ & $\delta^{(i)}$ & $\sigma^{(i)}$ \\
            \midrule
    \multirow{3}{*}{1} & 1 & $0.39 \pm 0.01$ & $1.24 \pm 0.22$ & $0.18 \pm 0.12$ \\ 
     & 2 & $0.35 \pm 0.02$ & $1.28 \pm 0.46$ & $2.34 \pm 0.44$ \\ 
     & 3 & $0.17 \pm 0.02$ & $7.31 \pm 9.09$ & $18.12 \pm 0.65$ \\ 
    \midrule
    \multirow{3}{*}{2} & 1 & $0.21 \pm 0.01$ & $0.00 \pm 0.33$ & $0.56 \pm 0.24$ \\ 
     & 2 & $0.31 \pm 0.02$ & $2.00 \pm 0.01$ & $2.67 \pm 0.02$ \\ 
     & 3 & $0.40 \pm 0.02$ & $14.79 \pm 0.65$ & $10.40 \pm 0.42$ \\ 
    \bottomrule
    \end{tabular}
\end{table*}

\begin{table}[ht]
    \centering
    \caption{Predictive performance metrics (on test set) after parameter estimation using Cochrane-Orcutt transformation for real-world watersheds. Evaluation metrics are computed on the original (untransformed) test set, as well as on the transformed test set.}
    \label{tab:realworld_perf}
    \begin{tabular}{ll|rrr}
           \toprule
            \multirow{2}{*}{watershed} & \multirow{2}{*}{$k$} & \multicolumn{3}{c}{metric} \\
            & &  $R^2$ & KGE & RMSE \\
           \midrule
           \multicolumn{5}{c}{original data} \\
           \midrule
    1 & 3 & 0.71 & 0.72 & 3.36 \\ 
    2 & 3 & 0.60 & 0.70 & 2.74\\  
    \midrule
    \multicolumn{5}{c}{transformed data} \\
    \midrule
    1 & 2 & 0.59 & 0.64 & 2.67 \\ 
    2 & 3 &  0.47 & 0.58 & 1.50\\ 
    \bottomrule
    \end{tabular}
\end{table}

The Gaussian SWR model selects $k=3$ windows for both Koksilah and Big Sur Rivers. Fig.~\ref{fig:real_world_kernels1} shows a clear dominance of the short-term history (a window close to a lag of 1 day) for the Koksilah River. Paired with the fact that only three highly overlapping windows were needed to model the catchment with high accuracy (Table \ref{tab:realworld_perf}), we can conclude that the Koksilah River is driven by simple hydrological processes. The third window is very wide with lower weight, and the first two windows have $\delta \simeq 1$, thus the combined kernel for watershed 1 resembles a truncated Student-$t$ distribution: the combined kernel remains unimodal, but the probability mass is more dispersed over the long-term lags compared to a Gaussian kernel. Moreover, a low lag value suggests a relatively simple catchment, which aligns with the observation that the Koksilah River is situated in a region characterized by extremely high precipitation and minimal aridity \citep{janssen2021hydrologic}. From previous works we can hypothesize that wet, non arid, catchments tend to be extremely simple with fast response times \citep{schoppa2020evaluating,spieler2020automatic}, which is strengthened by our experimental results. Since the bulk of precipitation transitions into streamflow in about 1 day, this catchment is likely dominated by shallow subsurface flow. Large amounts of precipitation in this catchment fully saturate and connect the shallow soil layers for much of the year, resulting in any excess precipitation pushing the water already in the soils towards the stream within one day. 

As we travel south along the western coast of North America towards the Big Sur River in Central California, aridity increases and total precipitation decreases slightly \citep{janssen2021hydrologic}. Although the Big Sur River is still considered a ``wet'' catchment, the Gaussian SWR model selects two more distinguishable peaks for watershed 2 compared to watershed 1, indicating an increase in catchment complexity in-part due to the increased aridity \citep{schoppa2020evaluating,spieler2020automatic}. The first large peak centred at $\delta=0$ indicates overland flow. This peak could be the result of the Big Sur River flowing through several developed towns where any precipitation falling on impermeable urban surfaces causes immediate drainage into the river. The wider second window could represent shallow subsurface flow in a similar fashion as watershed 1. The last window for the Big Sur River indicates that there is substantial baseflow in the catchment. With some precipitation flowing deeper underground towards less porous material, this partition of water will respond more slowly to water input compared to shallow subsurface flow and overland flow \citep{janssen2023learning}.

\begin{figure*}[ht]
    \centering
    \begin{subfigure}{0.7\textwidth}
        \centering
        \includegraphics[width=\textwidth]{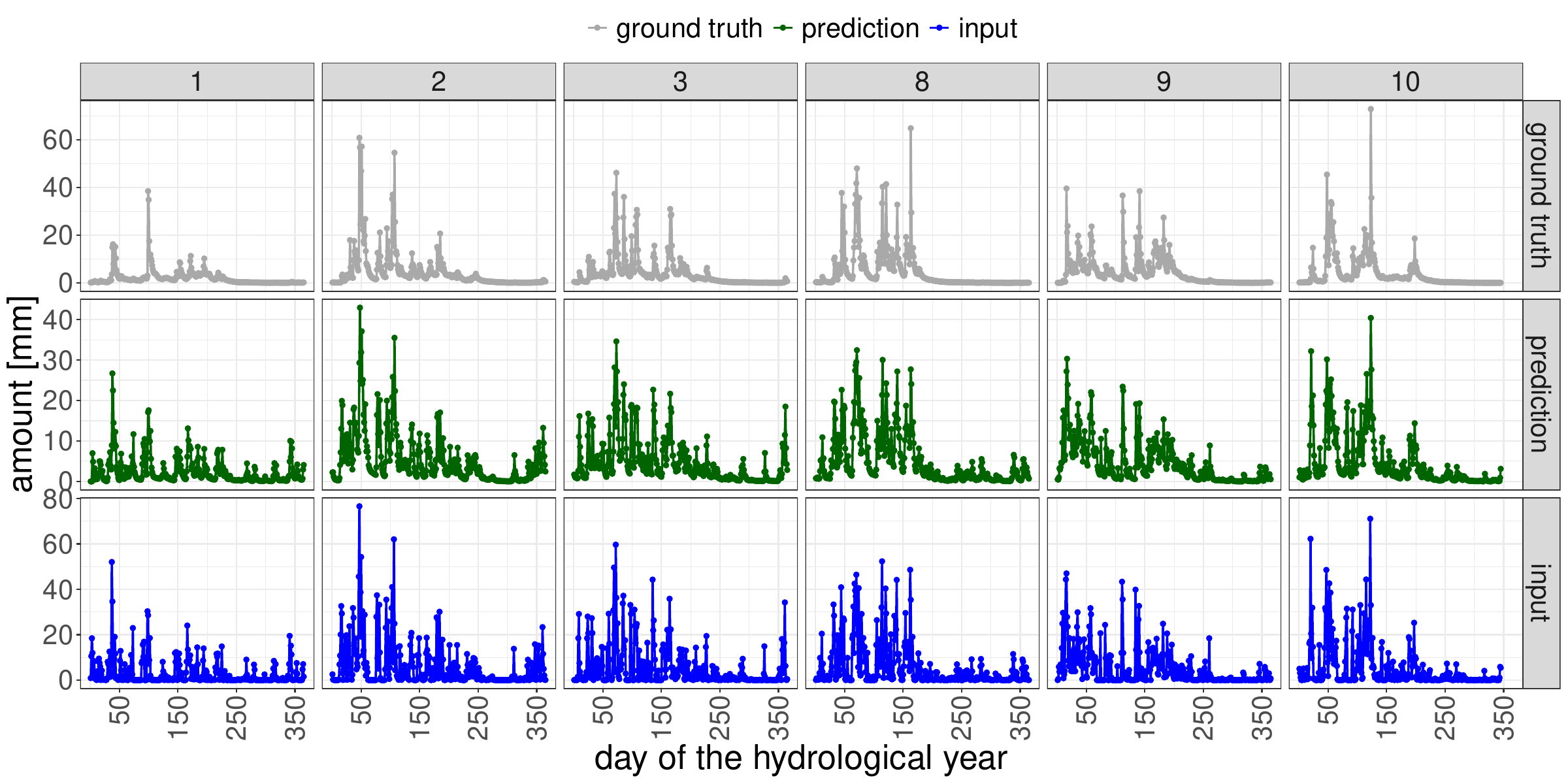}
        \caption{Watershed 1.}
        \label{fig:real_world_predictions1}
    \end{subfigure}
    \begin{subfigure}{0.7\textwidth}
        \centering
        \includegraphics[width=\textwidth]{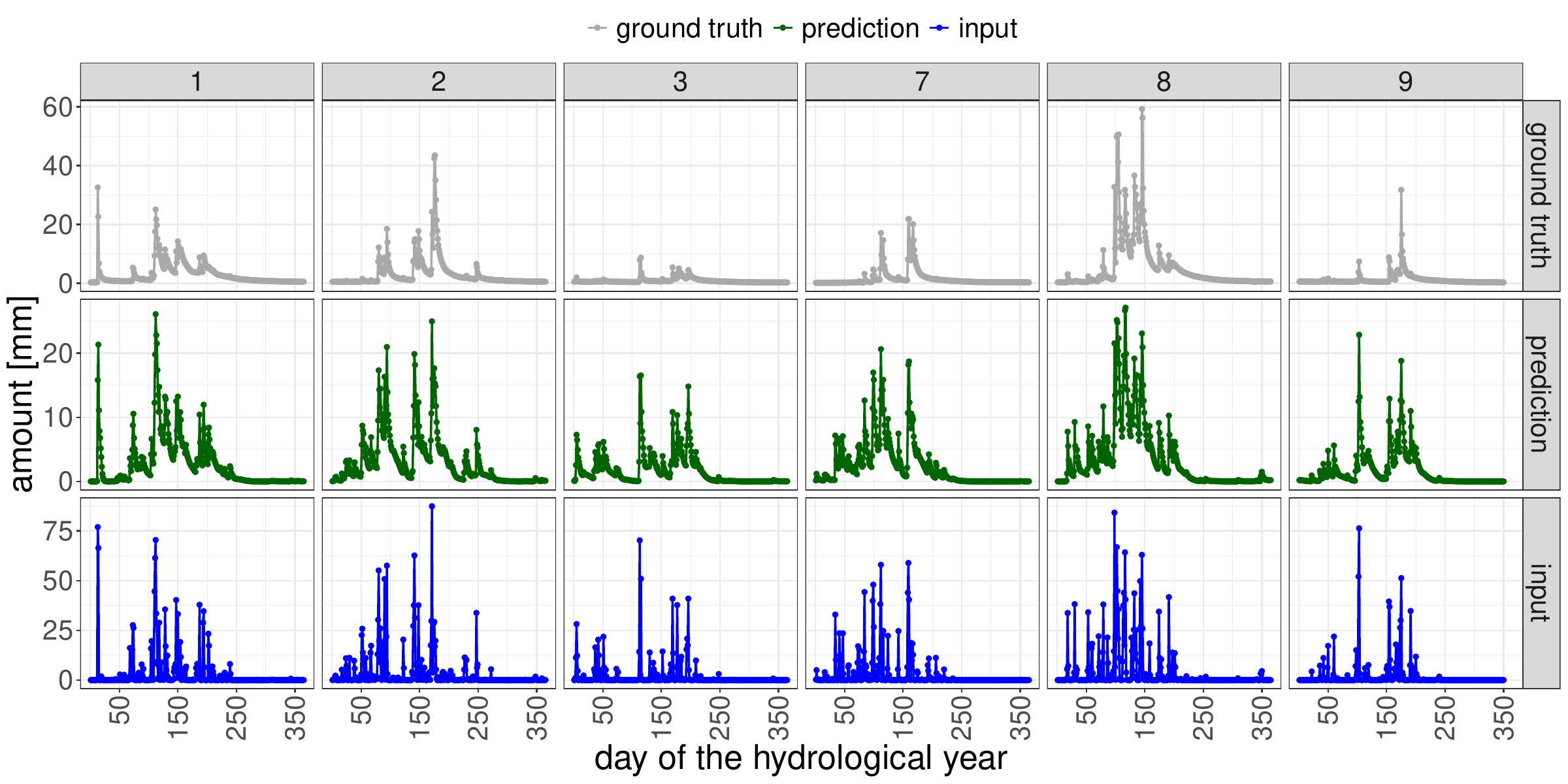}
        \caption{Watershed 2.}
        \label{fig:real_world_predictions2}
    \end{subfigure}
    \caption{Observed streamflow, predicted streamflow, and rainfall in the real-world experiments for six years of the test set}
    \label{fig:real_world_predictions}
\end{figure*}

Fig.~\ref{fig:real_world_predictions} compares the time series of predicted and observed waterflows for each watershed, along with the respective rainfall time series (all relating to untransformed data). To save space, the plots restrict the time domain to the first three and last three hydrological years of the test set. Since the Gaussian SWR model acts as a smoothing operator on the input time series, the prediction time series is less spiked than the rainfall input and generally matches the shape of the ground truth well. Particularly for watershed 1, time intervals with flat gauge values appear more noisy in the predictions. This becomes visible in the dry periods at the end of each hydrological year (time interval $[250, 365]$), where the ground truth response is almost exactly $0$ throughout, while the predictions are modestly positive.

To investigate the effect of sample size, we repeated training for Watershed 1 but with only the first six years of data, i.e., about 20\% of the previous training data comprising 29 years. The previously found third, very wide window with smaller weight seen in Figure~\ref{fig:real_world_kernels}(a) was no longer detected by BIC using the smaller dataset, but the combined kernel from the two windows closely resembled that shown in Figure~\ref{fig:real_world_kernels}(a) from three windows.

\subsection{Runtime Analysis}

All real world experiments were performed on an Intel Core i5 3GHz CPU with 8GB RAM running macOS Sonoma 14.4, where Algorithm~\ref{alg:training} was implementated in R. The runtimes in Table~\ref{tab:runtimes} are averages over 5 independent model fits to account for variations in the computation time. In each run, the Gaussian SWR model is trained for one real-world watershed using the same experimental setup as in the real-world study; time to compute predictions is excluded.  The maximum number of windows $k_{\max}$ is clearly important, but all runtimes are moderate.

\begin{table}
    \centering
    \caption{Runtimes of the Gaussian SWR model with $k_{\max} \in \{1, 2, 3\}$, averaged over 5 runs.}
    \label{tab:runtimes}
    \begin{tabular}{l r}
        \toprule
        $k_{\max}$ & runtime [s] \\
        \midrule
        1 & 12.4 \\
        2 & 75.6 \\
        3 & 608.8 \\
        \bottomrule
    \end{tabular}
\end{table}

\section{Discussion and Conclusion}
Our experiments validated the Gaussian SWR model in a controlled experimental environment and real-world scenarios.  Good predictive accuracy on a test set, as measured by $R^2$ for instance, was achieved throughout. The simulations demonstrated that achievable accuracy was dominated by the noise level, and rather independent of the number, position, size, and weighting of the ground truth windows. Moreover, multiple kernels were usually identifiable, with accurate parameter estimates. 
Thus, the proposed Gaussian SWR model has promising utility for hydrological inference.

Parameterizing the kernels as having Gaussian densities is easily extended to more flexible shapes, such as Student's $t$, Gamma, or asymmetric generalized Laplace distributions.  The latter two suggestions would enable long-tailed lag effects.

Relative to the more general DLM, the predictive performance of the Gaussian SWR model is comparable on the datasets used in our real-world experiments. When varying the number of time lags $q$, the DLM implementation by~\citet{demirhan2020dlagm} achieves test-set $R^2$ scores of 0.7 ($q=5$), 0.72 ($q=10$ and $q=20$) for watershed 1 and 0.51 ($q=5$), 0.55 ($q=10$), 0.58 ($q=20$) for watershed 2. Thus, the parametric assumption taken by Gaussian SWR, i.e. restricting the kernel shape to a mixture of Gaussians, does not significantly affect the predictive performance. However, the favourable parametrization employed by the Gaussian SWR model allows for straightforward interpretations of the model parameters: the window position and shape parameters are directly transferable to the concept of distinct flow paths in hydrology and deliver information on the peak lag of the runoff and its celerity distribution, respectively. Regression parameters $\bm{\beta}$ give indications of the relative importance of each flow path: while a dominant window with short lag typically models a strong short-term (overland) flow, well-distinguished windows with longer time lags indicate the presence of distinct long-term effects such as sub-surface or groundwater flows. From a statistical perspective, kernels representing long-term effects may approximate a constant line, thus adding an intercept could be a valid option for the proposed model --- for the hydrological interpretation, however, a large (slow-flow) window acts as a more accurate parametrization than a model with intercept.

From a hydrological viewpoint, a major limitation of the Gaussian SWR model is that the  kernels, particularly the parameters $\bm{\beta}$ and $\bm{\delta}$, do not change dynamically over the hydrologic year. The hydrologic year in many watersheds is characterized by drier and wetter periods, with distinct patterns in rainfall-runoff relationships. Related works by \citet{janssen2023learning} on the same watershed 1 demonstrate that the predictive performance can be increased to approximately $R^2=0.8$ using a functional data analysis approach with dynamic regression weights and sparsity. Potentially, the Gaussian SWR model setup could be extended in future work to account for such dynamic parameters at the cost of a higher number of model parameters and a more complex training procedure, while keeping its highly interpretable ability to separate flow paths.

The presence of significant precipitation falling as snow will impact runoff dynamics in colder regions. As a consequence, runoff lags and magnitudes of the distinct flows paths may vary significantly over the year, which is not covered by linear constant-lag models. To overcome this issue for watersheds with significant snow to rainfall ratios, an extension of the proposed Gaussian SWR model to account for non-linearities would further be of interest. We consider this as another potential future extension of the proposed model.

\section*{Acknowledgement}
This research was funded by the Canadian Statistical Sciences Institute (CANSSI) as a Collaborative Research Team project.
Stefan Schrunner and Anna Jenul gratefully acknowledge financial support from the Norwegian University of Life Sciences (project number 1211130114) to visit the University of British Columbia.  
Computing was supported in part by the Digital Research Alliance of Canada (alliancecan.ca).

Further, we would like to thank Asad Haris for contributing to the preparation of the dataset, as well as Oliver Meng and Junsong Tang for their work on graphical user interfaces to visualize the proposed model.

\bibliographystyle{abbrvnat}
\bibliography{literature}

\end{document}